\documentclass[final,12pt]{elsarticle}

\usepackage{subcaption}

\usepackage{amsmath}
\usepackage{amssymb}

\usepackage{cancel}

\usepackage{physics}

\usepackage{siunitx}

\usepackage[unicode=true]{hyperref}

\hypersetup{%
	plainpages=false,
	bookmarksopen=true,
	bookmarksnumbered=true,
	breaklinks=true,
}

\usepackage{amsthm}

\usepackage[capitalize,nameinlink,noabbrev]{cleveref}

\newtheoremstyle{break} 
  {}          
  {}          
  {}          
  {}          
  {\bfseries} 
  {.}         
  {\newline}  
  {}          
\theoremstyle{break}

\newtheorem{theoremx}{Theorem}[section]
\newenvironment{theorem}
  {\pushQED{\qed}\theoremx}
  {\popQED\endtheoremx}

\newtheorem{definitionx}[theoremx]{Definition}
\newenvironment{definition}
  {\pushQED{\qed}\definitionx}
  {\popQED\enddefinitionx}

\newtheorem{examplex}[theoremx]{Example}
\newenvironment{example}
  {\pushQED{\qed}\examplex}
  {\popQED\endexamplex}

\newtheorem{remarkx}[theoremx]{Remark}
\newenvironment{remark}
  {\pushQED{\qed}\remarkx}
  {\popQED\endremarkx}

\newtheorem*{remarkxx}{Remark}
\newenvironment{remark*}
  {\pushQED{\qed}\remarkxx}
  {\popQED\endremarkxx}

\usepackage{booktabs} 

\graphicspath{{figs/}}

\journal{arXiv}

\hypersetup{%
	pdfauthor={%
		Markus Lohmayer
		and Paul Kotyczka
		and Sigrid Leyendecker
	},
	pdftitle={Exergetic Port-Hamiltonian Systems: Modelling Basics}
}

\begin{document}

\begin{frontmatter}
	\title{Exergetic Port-Hamiltonian Systems: Modelling Basics}

	\author[fau]{Markus Lohmayer}
	\ead{markus.lohmayer@fau.de}

	\author[tum]{Paul Kotyczka}

	\author[fau]{Sigrid Leyendecker}

	\address[fau]{%
		Institute of Applied Dynamics,
		University of Erlangen-Nuremberg,
		Germany
	}

	\address[tum]{%
		Chair of Automatic Control,
		Technical University of Munich,
		Germany
	}

	\begin{abstract} 
		Port-Hamiltonian systems theory provides
		a structured approach to
		modelling, optimization and control
		of multiphysical systems.
		Yet, its relationship to thermodynamics seems to be unclear.
		The Hamiltonian is traditionally thought of as energy,
		although its meaning is exergy.
		This insight yields benefits:
		1. Links to the GENERIC structure are identified,
		making it relatively easy to borrow ideas
		from a popular framework for nonequilibrium thermodynamics.
		2. The port-Hamiltonian structure
		combined with a suitable bond-graph syntax
		is expected to become a main ingredient in
		thermodynamic optimization methods
		akin to exergy analysis and beyond.
		The intuitive nature of exergy and diagrammatic language
		facilitates interdisciplinary communication
		that is necessary for implementing
		sustainable energy systems and processes.
		Port-Hamiltonian systems are cyclo-passive,
		meaning that a power-balance equation
		immediately follows from their definition.
		For exergetic port-Hamiltonian systems,
		cyclo-passivity is synonymous with degradation of energy
		and follows from
		the first and the second law of thermodynamics
		being encoded as structural properties.
	\end{abstract}

\end{frontmatter}

\section{Introduction}%
\label{sec:introduction}

\subsection{Energy versus exergy}%
\label{ssec:energy_vs_exergy}

Energy is the most famous conserved quantity
and serves as a lingua franca throughout physics and beyond.
However, analysis of technical systems
based on the first law of thermodynamics alone (energy analysis)
is often not helpful or even misleading
because the quality of the energy
that is exchanged between components or subsystems
is not taken into account~\cite{2005Szargut}.
For instance, \SI{100}{\W} of heating power
can be obtained from \SI{100}{\W} of electric power
but the same heating power cannot be used
to again generate \SI{100}{\W} of electric power,
even if an ideal engine
without losses is assumed.
This is due to the second law of thermodynamics
which states that entropy (microscopic disorder)
can only be produced, never destroyed.
Because of irreversible degradation,
energy should not be regarded as a resource,
given that resource means
having potential to cause change
that is observable on the macroscopic level.

In contrast, exergy~\cite{1956Rant}
(also referred to as
available energy (of body and medium)~%
\cite{1873Gibbs,2002GaggioliRichardsonBowman,2002GaggioliPaulus}
and availability~\cite{1951Keenan})
takes energy quality into account.
For instance, \SI{100}{\W} of electric power
can be fully utilized to do work,
assuming and ideal engine.
Thus, electric power can be understood as
an energy rate (energetic power)
as well as an exergy rate (exergetic power).
In contrast to that, the amount of work
which can be obtained from a \SI{100}{\W} heat source
assuming again an ideal engine
is bounded by the Carnot efficiency
and depends not only on the source temperature
but also on the environment temperature.
If the source has a temperature of \SI{375}{\K} and
the environment has a temperature of \SI{300}{\K},
then the Carnot efficiency is
$(\SI{375}{\K} - \SI{300}{\K}) / \SI{375}{\K} = 0.2$.
Consequently, the energetic power of \SI{100}{\W}
amounts to \SI{20}{\W} of exergetic power,
meaning that no more than \SI{20}{\W} worth of work
can be generated from it in the given environment.
If this heat source would be an electric heater
then its exergy destruction rate
would consequently be \SI{80}{\W}.

\subsection{Exergy analysis and thermodynamic optimization}%
\label{ssec:exergy_analysis_and_thermodynamic_optimization}

Exergy destruction rates
(also called irreversibility rates)
are instrumental for thermodynamic optimization~%
\cite{1980Gaggioli,1985Kotas,1996Bejan,2005Szargut}.
Loss of exergy is proportional to production of entropy
with the proportionality factor being the environment temperature.
Hence, the Exergy Analysis method (in its original form),
compares the system under study
to its reversible counterpart.
In particular for heat engines,
operation in the reversible limit implies
infinitesimal heat transfer rates
and thus zero power.
Consequently, the Carnot efficiency
is a limit which is never attained in applications.

This mismatch between practice
and the well-established quasistatic reversible theory
prompted the emergence of
{Finite-Time Thermodynamics} (FTT) in the mid-1970s,
see~\cite{2011Andresen} for a review.
The central goal of FTT is to
establish more realistic performance limits
for the operation of systems featuring irreversible processes
under finite-time constraints.
In this case,
different objective functions generally lead to different optima.
However, all interesting objectives
make some trade-off between
minimizing exergetic losses (reversible limit)
and maximizing power (`free-fuel limit').
In the big picture, the latter extreme pays off only
when utilizing exergy that otherwise would be lost
(e.g.~solar energy and low-temperature heat).
Regarding consumption of carbon-based resources,
ecological criteria are of utmost relevance.
The FTT literature
is almost completely confined to
simple (endoreversible) models~\cite{1979aRubin,1979bRubin}
that consider only very few
irreversible processes.
In this way, performance bounds
are computed for different ecological and economical objectives
using variational methods
and in particular (averaged) optimal control theory.
This has led to general insight and principles~%
\cite{2001SalamonNultonSiragusaAndersenLimon}
but a considerable gap
between theory and applications still remains.

\subsection{Purpose of this research}%
\label{ssec:purpose_of_this_research}

Human energy systems and industrial processes
urgently need to achieve higher total efficiencies
while shifting dependencies to renewable resources.
This requires
integration of different energy domains
and a high-level of interconnection (with prosumers).
Ultimately, we need to deal with
quite complex networks
whose nonlinear transient dynamics are crucial.
In fact, this holds true for many applications,
from sustainable heat engine technologies
to large-scale district heating networks.
Thus, the development of a practical framework
to support engineering efforts in these directions
is of great importance.
In particular due to their compositional nature,
exergetic port-Hamiltonian systems
provide a solid foundation
for optimization- and control-oriented modelling
of energy systems and processes.
Their diagrammatic language helps
to formulate, understand and communicate
models and optimization goals
in interdisciplinary environments.
In~\cite{1992GovernToole1},
the use of `bond graph type of diagrams'
has been suggested as an alternative to Grassmann diagrams
which are commonly used as a visualization tool
for (steady-state) exergy analysis.
Linking diagrammatic expressions directly to
the physical and mathematical structure
of the underlying models
makes exergetic port-Hamiltonian systems
a powerful tool for (transient) exergy analysis
and related thermodynamic optimization methods.
Multi-energy systems,
regenerative thermal engines and heat pumps,
as well as buildings
are some interesting application areas
which could benefit from this research.

\subsection{Port-Hamiltonian systems and thermodynamics}%
\label{ssec:phs_and_thermodynamics}

In traditional systems theory,
building blocks interact
by exchanging arbitrary signals.
In contrast to that,
the essence of port-Hamiltonian systems theory
is to endow models of physical systems with a geometric structure,
called Dirac structure~\cite{%
1990Courant,
1993Dorfman,
1998DalsmoSchaft,
2009Merker,
2018BarberoCendraGarciaDiego},
that expresses the exchange of power
among system components
and possibly across system boundaries.
The central structural property
of a port-Hamiltonian (sub)system
is a power balance equation
which relates the stored power,
the dissipated power and the supplied power.
By definition, the dissipated power is always non-negative,
and consequently, the stored power is always
less than or equal to the supplied power.
This property is referred to as cyclo-passivity
(or cyclo-dissipativity).
If the storage function (Hamiltonian)
is bounded from below (implying finite storage capacity),
the system is said to be passive
(or dissipative).
If the dissipated power is always zero,
the system is said to be (cyclo-)lossless,
see~\cite{1972Willems,2017Schaft,2020Schaft}.

As in classical Hamiltonian mechanics,
the storage function of a port-Hamiltonian system
is traditionally thought of as an `energy' function.
Hence, the power balance equation should be of energetic type
and `dissipated power' should refer to the rate
at which `energy' is lost
due to phenomena such as
mechanical damping or electrical resistance.
Such use of language is clearly at odds
with the first law of thermodynamics.

Many popular applications of port-Hamiltonian systems
are confined to the electro-mechanical realm
where internal energy
(a macroscopic abstraction of mechanical energy at the microscopic level)
does not affect the dynamics of interest.
In~\cite{2009DuindamMacchelliStramigioliBruyninckx}
it is stated that
for isothermal%
\footnote{An isothermal system has homogeneous and constant temperature.}
systems
the Hamiltonian represents `free energy'
which can be lost.
Indeed, in the isothermal case,
the three concepts
Helmholtz free energy, Gibbs free energy and exergy
are closely related.

Following inspiration from bond-graph modelling,
previous attempts to
include thermal phenomena in the port-Hamiltonian framework
relied on non-linear power-continuous transformers.
In~\cite{2004EberardMaschke}, a port-Hamiltonian model
of thermal conduction in a solid is presented.
Entropy is used as a state variable
and the internal energy is accounted for in the energetic Hamiltonian.
Heat conduction is understood as
a power-continuous energy transformation process
which produces entropy.
According to the first law of thermodynamics,
this must lead to lossless systems.
In this case,
the port-Hamiltonian structure does not encode
that the dynamics is severely constrained
by the second law of thermodynamics.
In other words,
degradation of energy
and its implication on stability
does not manifest in the port-Hamiltonian structure.

A source of possible confusion
is that the `Hamiltonian'
of a dissipative port-Hamiltonian system
not only generates a Hamiltonian dynamics
but also a dissipative gradient dynamics.
In thermodynamics,
dissipation is synonymous with entropy production.
Therefore, it is not surprising that
entropy appears next to energy in the exergetic Hamiltonian.
For exergetic port-Hamiltonian systems,
the systems-theoretic meaning~\cite{1972Willems}
agrees with the thermodynamic meaning.

\subsection{Related frameworks}%
\label{ssec:related_frameworks}

Later attempts to
properly unify port-Hamiltonian systems
with thermodynamics diverged
into three distinct frameworks:
Firstly, just like exergetic port-Hamiltonian systems,
{Irreversible Port-Hamiltonian Systems}~%
\cite{2013RamirezMaschkeSbarbaro} use
the extensive thermodynamic variables as state variables
but their structure is significantly different.
The modification is necessary to encode
not only the first but also the second law of thermodynamics
while sticking with the total energy
as the Hamiltonian function.
Secondly, contact geometry is a natural setting
for thinking about Legendre transformations
which has been used in equilibrium thermodynamics
since~\cite{1973Hermann}.
The contact-geometric approach has been extended
to nonequilibrium thermodynamics
and open systems, see e.g.~%
\cite{2007EberardMaschkeSchaft}.
The core idea is to enlarge the state space
such that it also includes the intensive variables.
The dynamics are then restricted
to a Lagrangian submanifold which
is generated by a thermodynamic potential
and thus expresses material properties.
For one and the same thermodynamic system,
there are two contact-geometric descriptions,
namely one where energy (or a Legendre transformation of it)
and one were entropy (or a Legendre transformation of it)
is used as the generating function of the Legendre submanifold.
Thirdly, {Port-Thermodynamic Systems}~\cite{2018SchaftMaschke}
are based on a symplectization
of the contact-geometric description.
By adding one more dimension to the state space,
energetic and entropic representations
can be expressed simultaneously as projectivizations.
A comparison of the advantages and (current) limitations
of the different frameworks
is missing in the literature
and is also beyond the scope of the present article.
Yet, the order in which we listed the three frameworks
reflects a trend of adding more geometric structure
and in the two latter cases also more redundant state variables.
While this may be advantageous for certain purposes,
it has drawbacks as well.
Successful application of a modelling framework
also critically depends on
how easily it can be picked up by practitioners.
Exergetic port-Hamiltonian systems
shine because of their relative simplicity
and their readily available diagrammatic language.
This fits
one of our main research goals,
namely to develop a framework
which can form an adequate basis
for various near-term engineering efforts
to tackle the sustainability crisis.

\subsection{The GENERIC framework}%
\label{ssec:generic}

In the 1980s, some researchers started to combine
reversible Hamiltonian dynamics
with dissipative gradient dynamics~%
\cite{1980DzyaloshinskiiVolovick,1984Grmela,1984Kaufman,1984Morrison}.
The resulting framework for nonequilibrium thermodynamics
has later been termed GENERIC,
an acronym for
General Equation for Non-Equilibrium Reversible-Irreversible Coupling~%
\cite{1997GrmelaOettinger,1997OettingerGrmela}.
After the appearance of many articles
and two monographs~\cite{2005Oettinger,2018PavelkaKlikaGrmela},
its active development continues.

Thermodynamic systems consist of
an extremely large number of constituents
and therefore can be seen at multiple scales.
At the microscopic scale,
their governing equations are widely-believed
to be invariant under time-reversal transformation~%
\cite{2014PavelkaKlikaGrmela}
and the Hamiltonian formalism
is a natural choice to express
the reversible energy exchange
between kinetic and potential energy domains
of the numerous constituents.
Despite of this microscopic reversibility,
the dynamics turn out to be biased
at a more macroscopic scale:
An isolated system relaxes
and thereby approaches its equilibrium state
which maximizes entropy.
In some sense, entropy arises because of
the uncertainty (incomplete information)
regarding the microscopic state.
Entropy production can be seen
from the information perspective
as a dynamic maximally-unbiased (MaxEnt) estimate
given only knowledge about mesoscopic/macroscopic quantities~%
\cite{1957Jaynes,2019KlikaPavelkaVagnerGrmela}.

The relaxation processes
can be modelled directly at a more macroscopic scale
as (generalized) gradient dynamics.
This requires three ingredients:
1. An adequate choice of state variables
to describe the system at the desired scale.
2. An entropy function
which tends to its constrained maximum
during the approach to equilibrium.
3. A dissipation potential
which yields the constitutive relations
describing the relaxation processes,
see~e.g.~\cite{2018Grmela}.
Gradient dynamics uses
quadratic potentials,
whereas generalized gradient dynamics
uses non-quadratic potentials,
see~\cite{2016MielkeRengerPeletier}
for a statistical motivation.
Gradient dynamics is essentially equivalent to
{Linear Irreversible Thermodynamics} (LIT)~%
\cite{2016MielkeRengerPeletier}.
In LIT,
thermodynamic fluxes (such as heat flux)
depend linearly on
thermodynamic forces (such as temperature differences).
However, the linear relations (such as Fourier's law)
may depend arbitrarily on the state
(like in the case of temperature-dependent thermal conductivity).
A large class of relaxation phenomena
(including irreversible transport phenomena)
can be modelled using gradient dynamics / LIT,
see for instance~\cite{1984GrootMazur}.
An exception are
thermodynamic systems
which are so far from equilibrium
that the concept of temperature loses its meaning.
In other words,
systems for which a local equilibrium assumption cannot be made.
Some of them can be modelled by the Boltzmann equation.
In this case,
the GENERIC formulation hinges on
a non-quadratic dissipation potential~\cite{1997GrmelaOettinger}.
Alternatively,
constitutive relations of irreversible processes
may be stated in the even more general quasi-linear form.
This amounts to the choice of
a symmetric, positive semidefinite linear operator
(called dissipation operator)
which may depend on the system's state
and on the differential of the entropy function
with respect to the state variables,
see~\cite{2013HuettnerSvendsen}.
The quasi-linear form
is equivalent to
(generalized) gradient dynamics
if the dissipation operator fulfils
an integrability condition~\cite{2018Grmela}.
If the operator does not depend
on the differential of the entropy function,
this condition is trivially satisfied
and the resulting relations
are essentially equivalent to LIT~\cite{2005Oettinger}.
Since we are going to use
internal energy as a thermodynamic potential
and consequently entropy as a state variable
(see in particular~\cref{ex:compartment}),
the differential of the entropy function is constant.
Thus, we may consider
gradient dynamics,
quasi-linear relations,
and LIT
as essentially equivalent.

According to~\cite{2014PavelkaKlikaGrmela},
the GENERIC fixes a splitting:
The Hamiltonian dynamics
have to be invariant under time-reversal transformation,
and they must conserve entropy.
The (generalized) gradient dynamics
may not be time-reversal invariant,
they must conserve energy,
and they must be dissipative.
Both contributions have to conserve mass and volume.
The GENERIC framework
guides the modelling process
and asserts thermodynamic consistency of evolution equations.
Some progress has been made to
derive structure-preserving integration methods~%
\cite{2020ShangOettinger}
and to extend the framework to open systems
using ideas from port-Hamiltonian theory~%
\cite{2006Oettinger,2018BadlyanZimmer}.

\subsection{Port-Hamiltonian systems and exergy}%
\label{ssec:phs_and_exergy}

It was realized in~\cite{1997YdstieAlonso}
that exergy can be used as a storage function\footnote{%
	The concept of storage function generalizes
	that of Lyapunov function to open systems,
	see~\cite{1972Willems}.%
}
for passivity-based control.
This idea has been picked up several times
in the literature on port-Hamiltonian systems,
see e.g.~%
\cite{2011HoangCouenneJallutGorrec,2018ZitteHamrounCouennePitault}.
However, exergy was not used as
the Hamiltonian generating the dynamics,
but as an additional quantity used for control design.

In~\cite{2018BadlyanMaschkeBeattieMehrmann}
it was shown how the GENERIC formulation
of a compressible fluid
can be rewritten as a port-Hamiltonian system
by using an exergy-like Hamiltonian
and by factorizing the dissipation operator.
The approach was used in~%
\cite{2020HauschildMarheinekeMehrmannMohringBadlyanReinSchmidt}
for modelling of district heating networks.
Since both the GENERIC and the port-Hamiltonian framework
combine Hamiltonian and gradient dynamics,
it is not too surprising that such a reformulation is possible.

\subsection{Contribution}%
\label{ssec:contribution}

The result in~\cite{2018BadlyanMaschkeBeattieMehrmann}
suggests that
the port-Hamiltonian framework
may be linked to the GENERIC
by using exergy as a Hamiltonian function.
We continue to investigate this idea more deeply.
In doing so,
we arrive at a physically sound interpretation of dissipativity
in the context of classical (i.e.~isothermal) port-Hamiltonian systems.
Furthermore, we start with the development of
a thermodynamic modelling framework:
Exergetic port-Hamiltonian systems
borrow from the rich thermodynamic theory of the GENERIC framework
and combine it with the port-Hamiltonian structure
that is well suited for
interconnection, optimization and control.
In contrast to the result in~\cite{2018BadlyanMaschkeBeattieMehrmann},
the framework does not rely
on the factorization of the dissipation operator in the GENERIC.
Instead, it is based on a refined definition of resistive structure
that is in agreement with thermodynamic theory.
Throughout, we showcase
the diagrammatic representation of exergetic port-Hamiltonian systems
based on a slightly adapted bond-graph syntax.

\subsection{Assumptions and current limitations}%
\label{ssec:assumtions_limiations}

The framework is inherently limited to systems
for which
the local equilibrium assumption
can be made.
It thus seems adequate to
limit ourselves to
quadratic dissipation potentials
and the perspective of {Linear Irreversible Thermodynamics}.

In this article,
we restrict ourselves to
the finite-dimensional (lumped-parameter) setting.
Further, the examples in this work
do not include systems with mass transfer or chemical reactions.
Despite making extensive use of a bond-graph syntax,
we defer its precise definition to later.

\subsection{Outline}%
\label{ssec:outline}

In~\cref{sec:fundamental_definitions} we state the relevant definitions.
In~\cref{sec:exergy} we elaborate on the physical meaning of exergy.
In~\cref{sec:isothermal_systems} we show that
the present framework seamlessly extends classical port-Hamiltonian theory
which (implicitly) assumes
equilibrium with an isothermal environment.
In~\cref{sec:nonisothermal_systems} we concern ourselves with
the modelling non-isothermal systems.
In~\cref{sec:conclusions} we state our conclusions.

\section*{Terminology and notation}%
\label{sec:notation}
\addcontentsline{toc}{section}{Terminology and notation}

We always use the word `energy' in the thermodynamic sense.
We use Latin letters for extensive quantities
and lowercase Greek letters for intensive quantities.
In particular, we use
$u$ for internal energy,
$s$ for entropy, $\theta$ for temperature,
$v$ for volume, $\pi$ for pressure,
$m$ for mass, and $\mu$ for chemical potential.
Uppercase $U$, $S$, $\ldots$ denote
corresponding potential functions.
We use $N$ for total mass
because $M$ is used for the dissipation operator in the GENERIC.
A system or process is called closed
if mass (of every type of atom) is constant.
It is called isochoric/isothermal/isobaric
if volume/temperature/pressure is constant.

For tensorial quantities,
we use (abstract) index notation with Einstein's convention:
Indices of contravariant slots are written as superscript
and indices of covariant slots are written as subscript.
Repeated indices (up-down pairs) imply contraction.
With $\mathcal{X}$ a smooth manifold,
$T \mathcal{X}$ denotes the tangent bundle and
$T^* \mathcal{X}$ the cotangent bundle over $\mathcal{X}$.
We write $\mathcal{A} \rightarrow \mathcal{X}$
for a general vector bundle with total space $\mathcal{A}$
and base space $\mathcal{X}$.
When the latter is clear from the context,
we just write $\mathcal{A}$.
For vector bundles $\mathcal{A} \rightarrow \mathcal{X}$
and $\mathcal{B} \rightarrow \mathcal{X}$,
$\mathcal{A} \oplus \mathcal{B}$ is the vector bundle over $\mathcal{X}$
where $\forall x \in \mathcal{X} \colon
{(\mathcal{A} \oplus \mathcal{B})}_x = \mathcal{A}_x \times \mathcal{B}_x$.
Given a contravariant $2$-tensor field
$L \in \Gamma( T \mathcal{X} \otimes T \mathcal{X} )$,
the sharp map
$L^\sharp \colon T^* \mathcal{X} \rightarrow T \mathcal{X}$
is the (curried) function defined by
$\forall \alpha, \, \beta \in \Gamma \big( T^* \mathcal{X} \big) \colon
(L^\sharp(\alpha))(\beta) = L(\alpha, \beta)$.
Dually, the flat map
$\omega^\flat \colon T \mathcal{X} \rightarrow T^* \mathcal{X}$
corresponding to a covariant $2$-tensor field
$\omega \in \Gamma( T^* \mathcal{X} \otimes T^* \mathcal{X} )$
is a bundle map form the tangent to the cotangent bundle.
Its name derives from the fact that
in index notation,
it lowers the up-index of a tangent vector $X^j$
into the down-index of the covector $\alpha_i = \omega_{ij} \, X^j$.

\section{Fundamental definitions}%
\label{sec:fundamental_definitions}

To streamline the following presentation,
in this section we state suitable definitions
of GENERIC and port-Hamiltonian systems
and their underlying geometric structures.

\subsection{Definitions related to the GENERIC framework}%

Symplectic structures
are quintessential in Hamiltonian mechanics.
Poisson structures are more general,
allowing a type of degeneracy
that encodes conserved quantities other than energy.
For details we refer to~\cite{1983Weinstein,1999MarsdenRatiu}.

\begin{definition}[Poisson structure]
	Let $\mathcal{X}$ be a state manifold.
	Let $f,g,h \in C^\infty (\mathcal{X})$
	be arbitrary smooth functions (observables) on $\mathcal{X}$.
	A Poisson structure on $\mathcal{X}$
	is a bilinear and antisymmetric map
	$\pb{\cdot}{\cdot} \colon
	C^\infty (\mathcal{X}) \times C^\infty (\mathcal{X})
	\rightarrow C^\infty (\mathcal{X})$
	called Poisson bracket,
	which fulfils the Jacobi identity
	$\pb{\pb{f}{g}}{h} = \pb{\pb{f}{h}}{g} + \pb{f}{\pb{g}{h}}$
	and the Leibniz rule
	$\pb{f \, g}{h} = f \, \pb{g}{h} + g \, \pb{f}{h}$.%
	\label{def:poisson_structure}
\end{definition}

A Poisson structure on $\mathcal{X}$
makes the $\mathbb{R}$-vector space
of smooth functions $C^\infty(\mathcal{X})$
into a $\mathbb{R}$-algebra.
This so-called Poisson algebra
is an abstract Lie algebra
since for some fixed $h \in C^\infty(\mathcal{X})$,
$X_h := \pb{\cdot}{h} \colon
C^\infty (\mathcal{X}) \rightarrow C^\infty (\mathcal{X})$
is a $\mathbb{R}$-derivation on this algebra
by virtue of the Jacobi identity, i.e.
$X_h(\pb{f}{g}) = \pb{X_h(f)}{g} + \pb{f}{X_h(g)}$.

Vector fields
are (isomorphic to)
derivations on the commutative $\mathbb{R}$-algebra
of smooth functions $C^\infty (\mathcal{X})$
with pointwise multiplication.
The Leibniz rule says that for some fixed $h$ we have
$X_h(f \cdot g) = f \cdot X_h(g) + X_h(f) \cdot g$
and thus it asserts that $X_h$ is a vector field.

The Leibniz rule also implies that
for some $f,g \in C^\infty(\mathcal{X})$,
their bracket $\pb{f}{g}$
depends only on the differentials
$df, dg \in \Gamma(T^* \mathcal{X})$.
It follows that
the Poisson bracket can be defined
in terms of an antisymmetric contravariant $2$-tensor field
$L \in \Gamma \big( \wedge^{2} T \mathcal{X} \big)$ like so:
$\left . \pb{f}{g} \right|_x =
\pdv{f}{x^i} \, L^{ij}(x) \, \pdv{g}{x^j}$.
In terms of the Poisson bivector (field) $L$,
the Jacobi identity can be expressed as
$L^{il} \, \pdv{L^{jk}}{x^l} +
 L^{jl} \, \pdv{L^{ki}}{x^l} +
 L^{kl} \, \pdv{L^{ij}}{x^l} = 0$.

\begin{definition}[Hamiltonian vector field]
	Let $\mathcal{X}$ be a state manifold and
	let $\pb{\cdot}{\cdot}$ be a Poisson structure on $\mathcal{X}$
	which is defined by a Poisson bivector $L$.
	Let $H \in C^\infty (\mathcal{X})$.
	Then, $X_H = \pb{\cdot}{H} = L^\sharp(dH)$
	is called the Hamiltonian vector field
	corresponding to the Hamiltonian (function) $H$.
\end{definition}

The Jacobi identity also implies that
the map $C^\infty(\mathcal{X}) \rightarrow \Gamma(T \mathcal{X})$,
$h \mapsto X_h$
is a Lie algebra antihomomorphism
from the Poisson algebra (of generating functions)
to the Jacobi-Lie algebra of (Hamiltonian) vector fields, i.e.
$\pb{f}{h} \mapsto \commutator{X_h}{X_f}$.

\begin{definition}[Hamiltonian system]
	A Hamiltonian system
	is a triple
	$\left( \mathcal{X}, \, \pb{\cdot}{\cdot}, \, H \right)$
	where
	$\mathcal{X}$ is a state manifold,
	$\pb{\cdot}{\cdot}$ is a Poisson structure on $\mathcal{X}$
	defined by a Poisson bivector
	$L \in \Gamma \big( \wedge^{2} T \mathcal{X} \big)$,
	and $H \in C^\infty(\mathcal{X})$ is a Hamiltonian function.
	An observable $f \in C^\infty (\mathcal{X})$
	evolves according to
	$\dot{f} = \pb{f}{H} = X_H(f)$.
	The state $x$
	evolves according to
	$\dot{x} = X_H$, i.e.
	$\dot{x}^i = L^{ij}(x) \, \pdv{H}{x^j}$.%
	\label{def:hamiltonian_system}
\end{definition}

Due to antisymmetry,
$\dot{H} = \pb{H}{H} = 0$,
i.e.~the Hamiltonian is conserved.
If $L^\sharp \colon T^* \mathcal{X} \rightarrow T \mathcal{X}$
is degenerate,
there exist distinguished observables
$C_i \in C^\infty (\mathcal{X})$
called Casimir functions
such that
for all $x \in \mathcal{X}$ we have
$L^{ij}(x) \, \pdv{C_i}{x^j} = 0$.
Any function of Casimirs
is a (dependent) Casimir.
In particular, this holds for
the Poisson bracket of two Casimirs.
Locally, the number of independent Casimirs
is equal to the dimension of the kernel of $L^\sharp(x)$.
Since $\dot{C}_i = \pb{C_i}{H} = 0$
for any generating function $H$,
these conserved quantities
are referred to as structural invariants.

\begin{example}[harmonic oscillator]
	Let $\mathcal{X} = \mathbb{R}^2 \ni x = \left( q, \, p \right)$.
	In the canonical position and momentum coordinates,
	the bivector
	\begin{equation*}
		L =
		\left[
			\begin{array}{rr}
				 0 & 1 \\
				-1 & 0
			\end{array}
		\right]
	\end{equation*}
	defines a constant Poisson structure on $\mathcal{X}$
	for which the Jacobi identity is trivially satisfied.
	The Hamiltonian
	$H \colon \mathcal{X} \rightarrow \mathbb{R}$,
	$H(x) =
	q^2 / \left( 2 \, c \right) +
	p^2 / \left( 2 \, m \right)$
	represents the system's energy.
	The constant $c$ is the compliance of the spring
	and $m$ is the mass.%
	\label{ex:harmonic_oscillator}
\end{example}

To define reversibility and irreversibility,
we consider
a transformation applied to the equations of motion,
which primarily reverses time
but also flips the sign of state quantities having odd parity.
The concept of even and odd parities is purely axiomatic.
It reflects our expectation that certain quantities,
like velocities and momenta, flip their sign/direction (odd parity)
when a recording is suddenly played backwards,
whereas most other quantities,
like energy, entropy, configuration, pressure and temperature
momentarily stay the same (even parity).

\begin{definition}[time-reversal transformation]
	Time-reversal transformation $\mathrm{TRT}$
	is defined as
	$\mathrm{TRT}(y) = \mathrm{P}(y) \, y$
	where $\mathrm{P}(y) = \pm 1$
	is the parity of quantity $y$.
	Hence, we have $\mathrm{TRT} \circ \mathrm{TRT} = \mathrm{id}$.
	For time derivatives,
	$\mathrm{TRT} \bigl( \dv{y}{t} \bigr) =
	-\dv{}{t^\prime} \bigl( \mathrm{TRT}(y) \bigr)$
	where $t^\prime$ is the reversed time.%
	\label{def:time_reversal_transformation}
\end{definition}

\begin{theorem}[reversibility of Hamiltonian dynamics]
	Evolution equations of a Hamiltonian system
	$\dot{x}^i = L^{ij} \, \pdv{H}{x^j}$
	are invariant under $\mathrm{TRT}$ (reversible)
	if and only if
	$-\mathrm{P}(x^i) = \mathrm{P}(L^{ij}) \, \mathrm{P}(x^j)$.%
	\label{thm:reversibility_of_hamiltonian_dynamics}
\end{theorem}

\begin{proof}
	The condition follows from requiring
	equality of
	the original evolution equation
	and the transformed one.
	Applying $\mathrm{TRT}$ to the left-hand side yields
	\begin{equation*}
		\mathrm{TRT} \biggl( \dv{x^i}{t} \biggr) =
		-\dv{}{t^\prime} \bigl( \mathrm{TRT}(x^i) \bigr) =
		-\dv{}{t^\prime} \bigl( \mathrm{P}(x^i) \, x^i \bigr) =
		-\dv{x^i}{t^\prime} \, \mathrm{P}(x^i)
	\end{equation*}
	since the parities are constant.
	Applying $\mathrm{TRT}$ to the right-hand side yields
	\begin{equation*}
		\mathrm{TRT} \biggl( L^{ij} \, \pdv{H}{x^j} \biggr) =
		\mathrm{P}(L^{ij}) \, L^{ij} \, \pdv{H}{\bigl( \mathrm{P}(x^j) \, x^j \bigr)} =
		L^{ij} \, \pdv{H}{x^j} \, \mathrm{P}(L^{ij}) \, \mathrm{P}(x^j)
		\,.
	\end{equation*}
	Thus, the transformed equations
	are equal to the original equations
	if and only if
	$-\mathrm{P}(x^i) = \mathrm{P}(L^{ij}) \, \mathrm{P}(x^j)$.
\end{proof}

Regarding~\cref{ex:harmonic_oscillator},
the condition holds because
for a constant $L$ all parities $\mathrm{P}(L^{ij})$ are even
and we have $\mathrm{P}(q) = 1$, $\mathrm{P}(p) = -1$.
For more details we refer to~\cite{2014PavelkaKlikaGrmela}.

\begin{definition}[gradient structure]
	Let $\mathcal{X}$ be a state manifold.
	A gradient structure on $\mathcal{X}$
	is a dissipation potential
	$\Psi \colon T^* \mathcal{X} \rightarrow \mathbb{R}$
	with
	$\Psi(x, x^*) = \frac{1}{2} \, x^*_i \, M^{ij}(x) \, x^*_j$
	where $M$ is a symmetric positive semidefinite tensor (field).%
	\label{def:gradient_structure}
\end{definition}

Thus, the dissipation potential $\Psi$
is quadratic and convex in $x^* \in T^*_x \mathcal{X}$.

\begin{definition}[gradient system]
	A gradient system
	is a triple
	$\left( \mathcal{X}, \, \Psi, \, S \right)$
	where
	$\mathcal{X}$ is the state space,
	$\Psi$ defines the gradient structure on $\mathcal{X}$,
	and $S \in C^\infty(\mathcal{X})$ is the entropy function.
	The state evolves according to
	$\dot{x} =
	\left. \pdv{\Psi}{z} \right|_{z = \pdv{S}{x}}$
	which means
	$\dot{x}^i =
	M^{ij}(x) \, \pdv{S}{x^j}$.
	Due to positive semidefiniteness of $M$,
	$\dot{S} \geq 0$,
	i.e.~the dynamics is dissipative.%
	\label{def:gradient_system}
\end{definition}

If the dissipation operator
$M^\sharp \colon T^* \mathcal{X} \rightarrow T \mathcal{X}$
is degenerate,
there exist conserved quantities
called `metric Casimirs'~\cite{2018Grmela}.
For details regarding (generalized) gradient dynamics
we refer to~%
\cite{2016MielkeRengerPeletier,2018PavelkaKlikaGrmela}
and references therein.

\begin{theorem}[irreversibility of gradient dynamics]
	Evolution equations of a gradient system
	$\dot{x}^i = M^{ij} \, \pdv{S}{x^j}$
	are strictly \textit{not} invariant under $\mathrm{TRT}$
	(i.e.~irreversible)
	if and only if
	$\mathrm{P}(x^i) = \mathrm{P}(M^{ij}) \, \mathrm{P}(x^j)$.%
	\label{thm:irreversibility_of_gradient_dynamics}
\end{theorem}

\begin{proof}
	The proof proceeds along the same lines
	as the proof of~\cref{thm:reversibility_of_hamiltonian_dynamics},
	except that inequality of
	the original and the transformed evolution equation
	is required.
\end{proof}

\begin{definition}[GENERIC system]
	A GENERIC system is a $5$-tuple
	$\left( \mathcal{X}, \, \pb{\cdot}{\cdot}, \, \Psi, \, E, \, S \right)$
	where
	$\mathcal{X}$ is the state space,
	$\pb{\cdot}{\cdot}$ defines the Poisson structure,
	$\Psi$ defines the gradient structure,
	$E \in C^\infty(\mathcal{X})$ is the energy function,
	and $S \in C^\infty(\mathcal{X})$ is the entropy function.
	$\left( \mathcal{X}, \, \pb{\cdot}{\cdot}, \, E \right)$
	is required to be reversible according to~%
	\cref{thm:reversibility_of_hamiltonian_dynamics}.
	$\left( \mathcal{X}, \, \Psi, \, S \right)$
	is required to be irreversible according to~%
	\cref{thm:irreversibility_of_gradient_dynamics}.
	The entropy function must be a symplectic Casimir,
	i.e.~$L^{ij} \, \pdv{S}{x^j} = 0$,
	making the Hamiltonian dynamics non-dissipative.
	The energy function must be a metric Casimir,
	i.e.~$M^{ij} \, \pdv{E}{x^j} = 0$,
	making the gradient dynamics energy-conserving.
	Further, total mass $N$ and volume $V$
	must be both symplectic and metric Casimirs.
	The state evolves according to
	$\dot{x}^i =
	L^{ij}(x) \, \pdv{E}{x^j} +
	M^{ij}(x) \, \pdv{S}{x^j}$.%
	\label{def:generic_system}
\end{definition}

\subsection{Definitions related to port-Hamiltonian systems}%

In the case of a Poisson bivector
$L \in \Gamma \big( \wedge^{2} T \mathcal{X} \big)$,
degeneracy of
$L^\sharp \colon T^* \mathcal{X} \rightarrow T \mathcal{X}$
is related to conserved quantities,
whereas in the case of a presymplectic form
$\omega \in \Gamma \big( \wedge^{2} T^* \mathcal{X} \big)$,
degeneracy of
$\omega^\flat \colon T \mathcal{X} \rightarrow T^* \mathcal{X}$
is related to algebraic constraints.
Dirac structures combine both directions/features,
see~\cite{1990Courant,2013Bursztyn}.
Further,
their definition
may involve vector bundles
more general than $T \mathcal{X} \oplus T^* \mathcal{X}$
allowing for interconnection of systems
in the port-Hamiltonian framework,
see~\cite{1998DalsmoSchaft,2007CerveraSchaftBanos,%
2011BatlleMassanaSimo,2018BarberoCendraGarciaDiego}.
Instead of referring to (components of)
tangent and cotangent vectors,
one speaks more generally of flows and efforts.
Dirac structures admit various representations.
We base the following definition
on a particular one,
namely the hybrid input-output representation~%
\cite{1998BlochCrouch,2014SchaftJeltsema}.

\begin{definition}[Dirac structure]
	Let $\mathcal{X}$ be a manifold.
	Let $\mathcal{F} \rightarrow \mathcal{X}$ be a vector bundle
	which may have (a subbundle of) $T \mathcal{X}$ as a subbundle
	and let $\mathcal{E}$ be the dual bundle of $\mathcal{F}$.
	A Dirac structure $\mathcal{D}$ on $\mathcal{F} \rightarrow \mathcal{X}$
	is a subbundle of $\mathcal{F} \oplus \mathcal{E}$
	admitting the following representation
	(after grouping components of $\mathcal{F}$ and $\mathcal{E}$):
	For every $x \in \mathcal{X}$,
	\begin{equation*}
		\mathcal{D}_x =
		\left\{
			\left(
				\left[
					\begin{array}{c}
						f^A \\
						\hline
						f^B
					\end{array}
				\right]
				, \,
				\left[
					\begin{array}{c}
						e_A \\
						\hline
						e_B
					\end{array}
				\right]
			\right)
			\in \mathcal{F}_x \times \mathcal{E}_x
			\: \mid \:
			\left(
				\left[
					\begin{array}{c}
						f^A \\
						\hline
						e_B
					\end{array}
				\right]
				\, = \,
				J(x) \,
				\left[
					\begin{array}{c}
						e_A \\
						\hline
						f^B
					\end{array}
				\right]
			\right)
		\right\}
	\end{equation*}
	with $J(x)$ a skew-symmetric linear map.%
	\label{def:dirac_structure}
\end{definition}

Compared to the kernel representation,
this representation is biased in the sense that
flows $f \in \mathcal{F}$
and efforts $e \in \mathcal{E}$
are grouped into
`inputs' $e_A$, $f^B$
and `outputs' $f^A$, $e_B$.
This makes it suitable for encoding computational causality,
see~\cref{rem:causal_strokes}.
Dirac structures model
a power-conserving interconnection of system components
since the net power $e_i \, f^i$
vanishes due to skew-symmetry of $J(x)$.
Integrability of Dirac structures
is discussed in~\cite{1990Courant,1998DalsmoSchaft,2013Bursztyn,2009Merker}.

\begin{definition}[resistive structure]
	Let $\mathcal{X}$ be a manifold.
	Let $\mathcal{F} \rightarrow \mathcal{X}$ be a vector bundle
	and let $\mathcal{E}$ be the dual bundle of $\mathcal{F}$.
	A resistive structure $\mathcal{R}$ on $\mathcal{F} \rightarrow \mathcal{X}$
	is a subbundle of $\mathcal{F} \oplus \mathcal{E}$
	admitting the following representation:
	For every $x \in \mathcal{X}$,
	\begin{equation*}
		\mathcal{R}_x =
		\left\{
			\left( f, \, e \right)
			\in \mathcal{F}_x \times \mathcal{E}_x
			\: \mid \:
			f^i = R^{ij}(x) \, e_j
		\right\}
	\end{equation*}
	with $R$
	a contravariant symmetric positive semidefinite $2$-tensor (field).
	\label{def:resistive_structure}
\end{definition}

Consequently,
the dissipated power $e_i \, f^i$
is always non-negative.
The definition can be generalized by
using a hybrid input-output or kernel representation
but it is not necessary for our purposes.

\begin{definition}[port-Hamiltonian system]
	A port-Hamiltonian system is a $6$-tuple
	$\left(
		\mathcal{X}, \, \mathcal{F}_R, \, \mathcal{F}_B,
		\mathcal{D}, \, \mathcal{R}, \, H
	\right)$
	where
	$\mathcal{X}$ is the state space,
	$\mathcal{F}_R \rightarrow \mathcal{X}$
	is the bundle of resistive flows,
	$\mathcal{F}_B \rightarrow \mathcal{X}$
	is the bundle of boundary flows,
	$\mathcal{D}$ is the Dirac structure
	on $T \mathcal{X} \oplus \mathcal{F}_R \oplus \mathcal{F}_B$,
	$\mathcal{R}$ is the resistive structure
	on $\mathcal{F}_R$,
	and $H \in C^\infty(\mathcal{X})$ is the Hamiltonian.
	For an isolated system (where $\mathcal{F}_B$ is the zero vector bundle),
	the dynamics is determined by
	$\left( \dot{x}, \, f^R, \, 0, \, \dd H, \, e_R, \, 0 \right)
	\in \mathcal{D}
	\: , \:
	\left( f^R, e_R \right)
	\in \mathcal{R}$.
\end{definition}

We call
$\left( f^S, \, e_S \right) = \left( \dot{x}, \, \dd H \right)$ the storage,
$\left( f^R, \, e_R \right)$ the resistive
and $\left( f^B, \, e_B \right)$ the boundary port (variables).
From $\left( f^S, \, f^R, \, f^B, \, e_S, \, e_R, \, e_B \right) \in \mathcal{D}$
it follows that
the stored power $e_S \, f^S$
plus
the dissipated power $e_R \, f^R \geq 0$
is equal to
the supplied power $-e_B \, f^B$ (cyclo-passivity).

For now, we skip a formal definition of
the composition of port-Hamiltonian systems
(see~\cite{2007CerveraSchaftBanos,2011BatlleMassanaSimo,%
2018BarberoCendraGarciaDiego}).
We also skip a formal definition of
the bond-graph syntax used in the following sections.

\section{Exergy and its physical meaning}%
\label{sec:exergy}

Work can universally and fully be turned into heat.
However, this does not hold for the reverse direction.
For biological life and engineering,
production of work is central.
Work is a form of energy exchange
which carries 100\% exergy content
because all work can do work,
according to Newton's third law, Kirchhoff's circuit laws, etc.
In a certain sense,
work is energy which is under our control,
because it is exchanged at `our' mesoscale.
The mental and computational models behind our engineered devices
are able to resolve all relevant degrees of freedom
which are involved in exchange of work.
The first (widely known) study of physical laws
which limit production of work
was by Carnot~\cite{1824Carnot}.
He observed that the passage of `caloric' (heat)
from a high to a low temperature level
allows the production of work.
His theory was developed further by Thomson (Kelvin).
Carnot's theory allowed Thomson
to introduce the concept of absolute temperature~%
\cite{1848Thomson}
which in turn allowed him and Joule to give a concrete expression
for the amount of work which can be produced by Carnot's ideal engine~%
\cite{1854JouleThomson}.
Soon after, Clausius formulated
the first law of thermodynamics as we know it today~%
\cite{1850Clausius}.
At roughly the same time,
Thomson also introduced the concept of available energy~%
\cite{1852Thomson}.
Some years later,
Clausius introduced the concept of entropy
to efficiently express the second law of thermodynamics~%
\cite{1856Clausius}.
Seventeen years later,
Gibbs was able to give a more concrete expression to
Thomson's concept of available energy~%
\cite{1873Gibbs}.
The concept was further developed by Keenan
(who called it availability)~\cite{1951Keenan}.
Rant gave the concept the name exergy~%
\cite{1956Rant}.
Since then, there has been active development of
thermodynamic design and optimization methods
based on and related to exergy (analysis),
see the engineering monographs~%
\cite{1980Gaggioli,1985Kotas,1995BejanTsatsaronisMoran,2005Szargut}.
With hindsight, we could say that
the idea which Carnot had about `caloric'
matured to eventually become the exergy concept.
For him, `caloric' was always conserved
because he conducted his study by imagining an ideal engine.

As a main takeaway,
it is crucial to distinguish between
reversible/nondissipative
and irreversible/dissipative processes.
Exergy is a thermodynamic quantity
which is conserved by the former
and destroyed by the latter.
We can divide exergy components into two kinds, namely
pure exergy components (kinetic, potential, magnetic, electric),
which can be exchanged as work
and
so-called physical exergy components (corresponding to internal energy),
which are involved in irreversible processes.

The upshot of Thomson's 1852 paper~\cite{1852Thomson}
introducing the available energy concept
was that eventually all available energy will be destroyed.
However, the boundary conditions and expansion rate of the universe
are not well known to mankind
and therefore we must not conclude that
the physical universe
approaches thermodynamic equilibrium~\cite{2013Wang}
which in this context is often referred to as
the dead state (of the system)
or heat death (of the universe).
This uncertainty
and the pessimism about life carried by this terminology
speak against its use.

Every engine has an underlying operational design
which allows it to extract part of its exergy input
and turn it into work,
according to the intent of its designers.
Similarly, biological life
can be understood as the interaction of
open thermodynamic systems
which function based on an `operational design' (DNA, etc.).
This is in stark contrast to other (umanaged) thermodynamic processes
in nature which merely happen spontaneously~\cite{2017Wang}.
Real machines and beings destroy exergy,
meaning that they cannot operate completely reversibly.
Some level of exergy destruction
is required
to meet robustness and performance requirements.

The first and second law of thermodynamics
are in principle not required to
model and simulate physical systems.
However, they serve important purposes:
On the one hand, they limit the set of
physically meaningful governing equations,
which provides structure and guidance
in the modelling process.
Indeed, all models expressible in
the GENERIC and the introduced framework
are coherent with the two laws.
On the other hand, they are of utmost relevance for
developing the operational design,
which is by no means less important than
the underlying physical laws,
at least for engineers.
The proposed framework informs the design process
by clearly indicating
how the theoretically available work
is lost or used by the system.

In the GENERIC and the present framework,
the distinction between reversible and irreversible processes
manifests itself in two types of relations,
namely Poisson/Dirac
and gradient/resistive structures.
In the GENERIC, each type has
an associated generating function:
The reversible dynamics is generated
by the (total) energy function
and the irreversible dynamics is generated
by the entropy function.
For exergetic port-Hamiltonian systems,
reversible and irreversible aspects
are combined into a single storage function.
This exergetic Hamiltonian is understood as
a multiphysical and systems-theoretical generalization
of what Gibbs called
the `available energy of body and medium'~\cite{1873Gibbs}.
The environment (medium) is an infinite reservoir.
It servers as a reference for assessing
the potential of the system (body) to do work
as it relaxes to equilibrium (with itself and the environment).
Since the environment is always in equilibrium (with itself),
its exergy content is zero by definition.
Thus, we can always consider the environment as part of the system
(without changing its Hamiltonian).
More concretely,
the environment can be understood as an atmosphere
whose temperature and pressure remain constant.
For many energy systems such as
regenerative heat engines
or district heating networks,
a typical weather condition
or ground temperature
serves as a natural reference.
In summary,
the exergy content of a system (including its environment) is
the amount of work that can be extracted
in the reversible limit
before the system reaches its equilibrium state
where its exergy content becomes zero.

\subsection{Reversible heat exchange and the Carnot engine}%
\label{ssec:carnot_engine}

The maximum amount of work
which can be extracted from heat
was studied by Carnot~\cite{1824Carnot}
using the concept of
an ideal heat engine,
see e.g.~\cite{1985Callen} (p.~118).
Since the engine executes a fully reversible cycle,
it generates no entropy
and achieves the highest efficiency
which is possible
for any heat engine operating
between the \textit{two} given thermal reservoirs.
\begin{figure}[!htb]
	\centering
	\includegraphics[width=\textwidth]{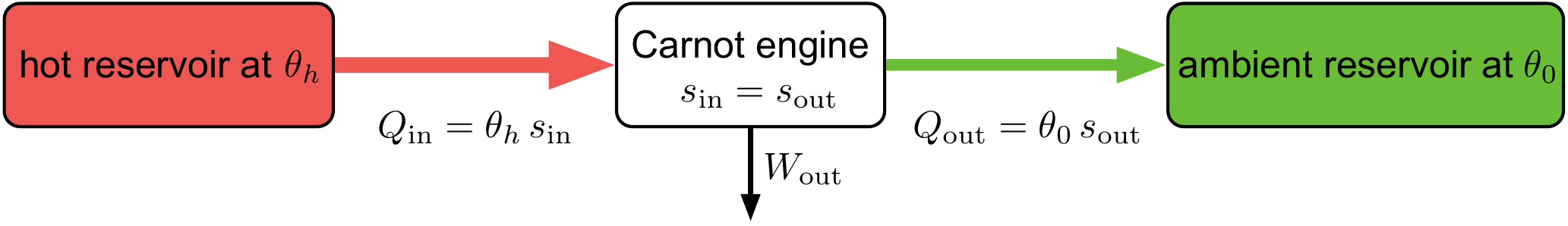}
	\caption{A Carnot engine operating between
		two isothermal reservoirs.
		In each cycle, the fictitious engine
		consumes the heat $Q_\text{in}$
		at the temperature level $\theta_h$ and
		rejects the heat $Q_\text{out}$
		at the temperature level $\theta_0$
		to perform the work $W_\text{out}$.
		Entropy is conserved.
	}%
	\label{fig:carnot_engine}
\end{figure}
\Cref{fig:carnot_engine} depicts a Carnot engine
extracting work
from a hot reservoir with constant temperature $\theta_h$.
The reversible heat intake per cycle
$Q_\text{in} > 0$
is associated with an entropy intake
$s_\text{in} = Q_\text{in} / \theta_h > 0$.
Cyclic operation implies that
entropy cannot accumulate in the engine.
Reversible operation implies that
no entropy is generated in the engine.
Consequently, the intake of entropy $s_\text{in}$
is balanced by
a discharge of entropy $s_\text{out} = s_\text{in}$.
This is associated with
a reversible discharge of heat
$Q_\text{out} = \theta_0 \, s_\text{out} > 0$
to the ambient reservoir at constant temperature $\theta_0$.

The entropy balance equation
$s_\text{out} = s_\text{in}$
can thus be written as
\begin{equation}
	\frac{1}{\theta_h} \, Q_\text{in}
	\: = \:
	\frac{1}{\theta_0} \, Q_\text{out}
	\,.
	\label{eq:carnot_entropy_balance}
\end{equation}
This yields the energy balance equation
\begin{equation}
	W_\text{out}
	\: = \:
	Q_\text{in}
	\: - \:
	Q_\text{out}
	\: \overset{\eqref{eq:carnot_entropy_balance}}{=} \:
	Q_\text{in}
	\: - \:
	\frac{\theta_0}{\theta_h} \, Q_\text{in}\\
	\: = \:
	\frac{\theta_h - \theta_0}{\theta_h} \, Q_\text{in}
	\,.
	\label{eq:carnot_energy_balance}
\end{equation}
The ratio
\begin{equation}
	\eta_\text{Carnot}
	\: = \:
	\frac{\theta_h - \theta_0}{\theta_h}
	\: < \: 1
	\label{eq:carnot_efficiency}
\end{equation}
is called Carnot efficiency
and expresses how much work $W_\text{out}$ can be obtained
for a given heat intake $Q_\text{in}$
in the reversible limit.
Consequently, we call
$\dot{W} = \eta_\text{Carnot} \: \dot{Q}$
the exergetic power of the heat source
and note that it is defined relative to
a fixed reference temperature $\theta_0$.
Heat exchanged at a higher temperature carries more exergy.

As an aside,
when irreversibility of thermal conduction is considered,
the efficiency in the opposite extreme
(maximum-work or `free-fuel' limit)
was studied via
the Curzon-Ahlborn engine~%
\cite{1975CurzonAhlborn,2007Miranda}.
Surprisingly, it turns out to be of a similar form, namely
$\eta_\text{CA} = 1 - \sqrt{\theta_0 / \theta_h}$.
For exergetic port-Hamiltonian systems,
the Carnot efficiency is immediately relevant,
since Dirac structures model
reversible exchange of power,
generalizing ideal wires from circuit theory.

\subsection{Exergetic storage function}%

Let us consider a closed and isochoric system
with entropy $s$
and internal energy $u$ given according to
its fundamental equation $u = U(s)$.
We assume that the system is in equilibrium with itself
but not necessarily with its environment.
The system's exergy $A$ is
\begin{equation}
	A(s)
	\: = \:
	U(s)
	\: - \:
	U(s_0)
	\: - \:
	\theta_0 \left( s - s_0 \right)
	\label{eq:exergy_s}
\end{equation}
where $s_0$ is its entropy
once the overall system has reached its equilibrium state.
Thus,
an infinitesimal change of exergy
is written as
$\dd A = \bigl( \theta(s) - \theta_0 \bigr) \, \dd s$
with $\theta(s) = \pdv{U(s)}{s}$.
Reversible exchange of heat power $\dot{Q}$
at the temperature level $\theta$
is associated with
an entropy exchange rate $\dot{s} = \frac{1}{\theta} \, \dot{Q}$.
Consequently, the corresponding exergetic power is given as
\begin{equation}
	\dot{A} = \bigl( \theta(s) - \theta_0 \bigr) \dot{s} =
	\frac{\theta - \theta_0}{\theta} \, \dot{Q}
	\,.
	\label{eq:exergy_carnot_efficiency}
\end{equation}
The exergetic power $\dot{A}$
is equal to the (energetic) thermal power $\dot{Q}$
multiplied with the Carnot efficiency
which is defined by
the temperature $\theta$
at which the thermal power is exchanged
and the environment temperature $\theta_0$,
see~\cref{eq:carnot_efficiency}.

Now, let us consider a closed system
with entropy $s$, volume $v$,
and internal energy $u$ given according to
its fundamental equation $u = U(s, v)$.
The system further has potential energy $E_\text{pot}$
depending on its configuration $q$
and kinetic energy $E_\text{kin}$ depending on its momentum $p$.
Its exergy $A$ is
\begin{multline}
	A(s, v, q, p)
	\: = \:
	U(s, v) - U(s_0, v_0)
	\: - \:
	\theta_0 \left( s - s_0 \right)
	\: + \:
	\pi_0 \left( v - v_0 \right)
	\: + \\ + \:
	E_\text{pot}(q) - E_\text{pot}(q_0)
	\: + \:
	E_\text{kin}(p)
	\,.
	\label{eq:exergy_svqp}
\end{multline}

The exergy function is obtained from the energy function
by adding
linear terms which determine the equilibrium
and constant terms which make the exergy zero at equilibrium.
The physical meaning of the term linear in $s$
has been explained above
using the Carnot engine as a theoretical device.
The meaning of the term linear in $v$
can be explained as follows:
The infinitesimal change of exergy
caused by an infinitesimal change of volume
is $\bigl( \pi(s, v) - \pi_0 \bigr) \, \dd v$
with pressure\footnote{%
	Pressure is defined with a minus sign because
	the system loses energy as it expands.
}
$\pi(s, v) = -\pdv{U(s,v)}{v}$.
If the system expands at the rate $\dot{v}$,
it has to displace the atmosphere
which is assumed to have a fixed pressure $\pi_0$.
This requires the mechanical power $\pi_0 \, \dot{v}$.
Hence, only
$\left( \pi - \pi_0 \right) \dot{v}$
remains as exergetic power.
The first line in~\cref{eq:exergy_svqp} represents
the physical exergy,
while the second line represents
the (macroscopic) mechanical energy/exergy.
The constant $-E_\text{pot}(q_0)$,
corresponding to the configuration with least potential energy,
makes the total exergy zero in the equilibrium state.

In the sequel,
we usually omit the constant terms in the exergy function
to save space
and because in practice these constants are often not known a priori.
In general, we consider potentials
differing only by an additive constant
to be equivalent.

\subsection{Two similar viewpoints}%

According to~\cref{def:generic_system},
the state of a GENERIC system evolves according to
$\dot{x}^i =
L^{ij}(x) \, \pdv{E}{x^j} +
M^{ij}(x) \, \pdv{S}{x^j}$.
Using the degeneracy conditions,
we can rewrite this as
\begin{subequations}
\begin{equation}
	\dot{x}^i \: = \:
	\left( \theta_0 \, L^{ij}(x) - M^{ij}(x) \right) \pdv{\Phi}{x^j}
	\label{eq:generic_phi_evolution}
\end{equation}
with a single generating function
$\Phi = -S
+ \frac{1}{\theta_0} \, E + \frac{\pi_0}{\theta_0} \, V
- \frac{\mu_0}{\theta_0} \, N$
where
$\frac{1}{\theta_0}, \,
 \frac{\pi_0}{\theta_0}, \,
 \frac{\mu_0}{\theta_0} \in \mathbb{R}^+$
are some constant multipliers.
The GENERIC models the approach to the equilibrium state $x_0$
and $\Phi$ is a Lyapunov function for the stability of $x_0$~%
\cite{1997GrmelaOettinger,2018PavelkaKlikaGrmela}.
Adding some constant shifts,
$\Phi$ can equivalently be defined as
\begin{equation}
	\Phi(x) = -S(x)
	+ \frac{1}{\theta_0} \bigl( E(x) - E(x_0) \bigr)
	+ \frac{\pi_0}{\theta_0} \bigl( V(x) - V(x_0) \bigr)
	- \frac{\mu_0}{\theta_0} \bigl( N(x) - N(x_0) \bigr)
	\,.
	\label{eq:generic_phi_potential}
\end{equation}%
\label{eq:generic_phi}
\end{subequations}
\Cref{eq:generic_phi_potential} shows
the relationship between GENERIC and the maximum entropy principle:
The equilibrium state $x_0$ maximizes the entropy $S$
under the constraints of
constant total energy $E$, volume $V$, and mass $N$.
Thus,
$\frac{1}{\theta_0}$,
$\frac{\pi_0}{\theta_0}$
and $\frac{\mu_0}{\theta_0}$
can be seen as Lagrange multipliers
for constraints stemming from fundamental conservation laws
that hold for isolated systems.
The multipliers are thus determined by
the values of the intensive variables at equilibrium,
as already suggested by the notation.

We can also write the GENERIC in the unconventional,
yet equivalent form
\begin{subequations}
\begin{equation}
	\dot{x}^i \: = \:
	\left( L^{ij}(x) - \frac{1}{\theta_0} M^{ij}(x) \right) \pdv{A}{x^j}
	\label{eq:generic_a_evolution}
\end{equation}
with generating function
\begin{equation}
	A(x) \: = \:
	E(x) - E(x_0)
	- \theta_0 \bigl( S(x) - S(x_0) \bigr)
	+ \pi_0 \bigl( V(x) - V(x_0) \bigr)
	- \mu_0 \bigl( N(x) - N(x_0) \bigr)
	\,.
	\label{eq:generic_a_potential}
\end{equation}%
\label{eq:generic_a}
\end{subequations}
While~\cref{eq:generic_phi}
corresponds to maximizing entropy subject to
constraints which must be satisfied for an isolated system,
\cref{eq:generic_a}
corresponds to minimizing the exergy function
defined by~\cref{eq:generic_a_potential}.
If we consider a system
consisting of a subsystem (body)
together with an infinitely large environment (medium),
\cref{eq:generic_a_potential}
may be seen as a Lagrangian
for the maximum amount of work $E(x) - E(x_0)$
that can be extracted from the overall system
while keeping its total entropy\footnote{
	The constant entropy constraint
	is of course related to the fact
	that we can only determine the maximum
	if we consider reversible extraction.
	In \textit{an} irreversible case,
	the answer depends
	not only on the laws of thermodynamics
	but also on the considered operational design
	and further constraints,
	such as finite speed, etc.
} $S$,
the volume $V$ and the mass $N$ constant.
Since the medium is infinitely large,
its intensive variables
are equal to the intensive variables
of the overall system at equilibrium.
These variables are
the Lagrange multipliers $\theta_0, \, \pi_0, \, \mu_0$.
Of course,~\cref{eq:generic_a_potential}
is also a Lyapunov function
for thermodynamic equilibrium.
This makes immediate sense
since, again, exergy is the amount of work
which can be extracted from the system
until it reaches thermodynamic equilibrium
where no (spontaneous) changes can occur~\cite{1951Keenan}.

We conclude that we are dealing with two similar viewpoints, namely
one which is biased towards entropy and relaxation
and another one which is biased towards energy and its degradation.
The former is more natural for the GENERIC framework,
while the latter is taken by
exergetic port-Hamiltonian systems.

\section{Isothermal systems}%
\label{sec:isothermal_systems}

In this section,
we introduce classical dissipative port-Hamiltonian systems
by means of the simplest example,
namely the damped harmonic oscillator.
After a discussion of
the physical meaning of the Hamiltonian function,
we will see how
a subtle modification of the oscillator model
leads us to an isothermal exergetic port-Hamiltonian system.
We conclude that
the present framework
is a straightforward extension of the classical theory.
The extension provides
a thermodynamic structure to port-Hamiltonian systems
which can be understood as
a compositional version of the GENERIC structure.

\subsection{Classical port-Hamiltonian systems}%

In the classical port-Hamiltonian framework,
physical systems are ultimately comprised of
energy storage components,
energy routing components,
and (free) energy dissipating components,
as the following example shows:%

\begin{example}[classical model of the damped harmonic oscillator]
	The free vibration of
	the damped oscillator
	in~\cref{fig:oscillator_sketch}
	is commonly modelled via
	the differential equation
	$m \, \ddot{q} + d \, \dot{q} + \frac{1}{c} \, q \, = \, 0$,
	where
	$q$ is the displacement,
	$m$ is the mass,
	$d$ is the damping coefficient,
	and $c$ is the spring compliance.

	\begin{figure}[!htb]
		\centering
		\begin{subfigure}{.2\textwidth}
			\centering
			\includegraphics[height=110pt]{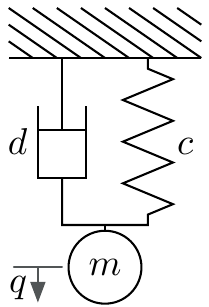}
			\caption{sketch}%
			\label{fig:oscillator_sketch}
		\end{subfigure}%
		\begin{subfigure}{.8\textwidth}
			\centering
			\includegraphics[height=110pt]{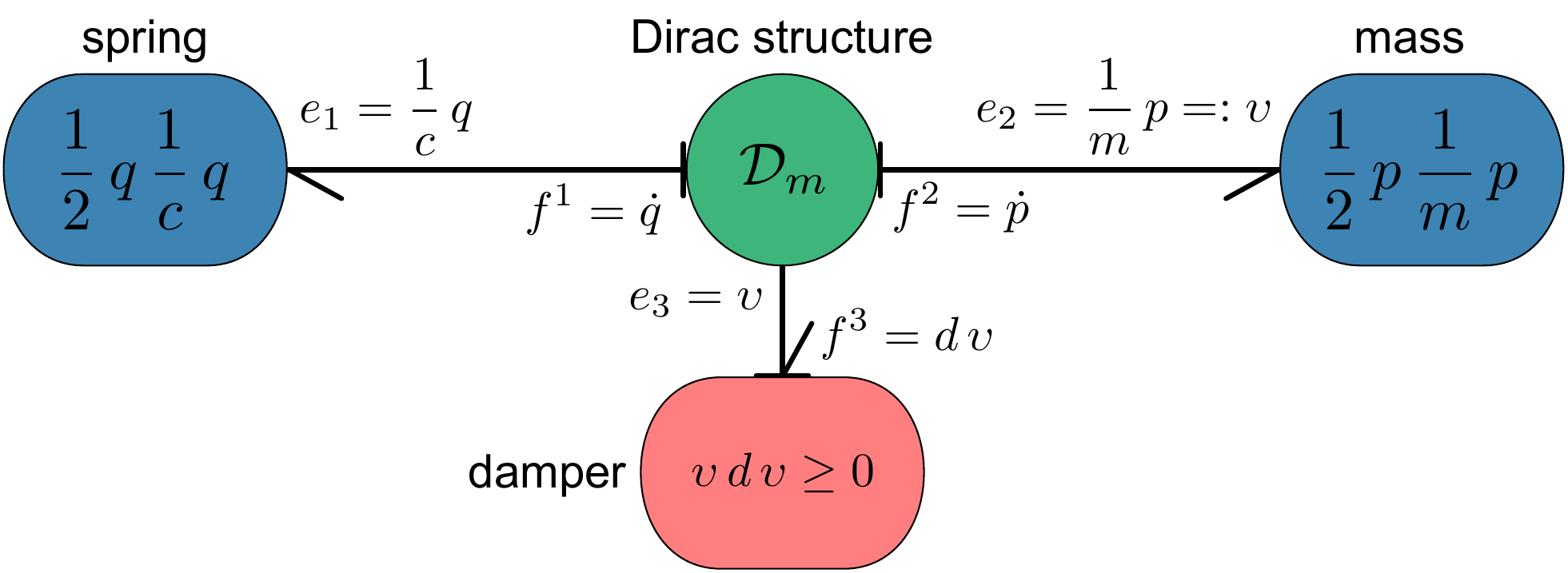}
			\caption{bond-graph expression}%
			\label{fig:oscillator_bond_graph}
		\end{subfigure}
		\caption{Damped harmonic oscillator.}%
		\label{fig:oscillator}
	\end{figure}

	The port-Hamiltonian formulation uses separate state variables
	for each storage component,
	namely the extension of the spring $q$
	and the momentum of the mass $p$.
	\Cref{fig:oscillator_bond_graph}
	shows a bond-graph expression of
	the port-Hamiltonian system.
	Regarding the syntax,
	blue boxes represent storage components,
	green boxes represent the Dirac structure,
	and red boxes represent the resistive structure.
	The sum of the terms annotated inside the blue boxes
	represents the Hamiltonian $H$.
	The term annotated inside the red box
	represents the dissipated power.
	An arrowhead on every bond indicates
	the direction in which power flows
	when the respective pairing
	$e_i \, f^i$ (fixed $i$)
	takes a positive value.
	We fix this direction
	such that positive values correspond to
	stored power, dissipated power, and power supplied to other systems.
	This will allow us to eventually omit the arrowheads
	resulting in clearer diagrams and simplified communication.
	For now, we also annotate each bond with a causal stroke:
	A transversely-oriented bar is placed on that end
	which assigns the value of the flow variable,
	see~\cref{rem:causal_strokes}
	at the end of the section.
	In this article,
	we additionally annotate each bond with formulaic expressions
	of its associated flow and effort variables.

	The state
	$x = \left( q, \, p \right)
	\in \mathcal{X} = \mathbb{R}^2$
	evolves according to
	\begin{subequations}
		\begin{align}
			\left( f^1, \, f^2, \, e_1, \, e_2 \right)
			\: &= \:
			\left(
				\dot{q}, \,
				\dot{p}, \,
				\frac{\partial H}{\partial q}(x), \,
				\frac{\partial H}{\partial p}(x)
			\right)
			\label{eq:oscillator_storage}
			\\[0.25em]
			\left[
				\begin{array}{c}
					f^1\\
					f^2\\
					\hline
					e_3
				\end{array}
			\right]
			\: &= \:
			\left[
				\begin{array}{rr|r}
					 0 & 1 &  0\\
					-1 & 0 & -1\\
					\hline
					 0 & 1 &  0
				\end{array}
			\right]
			\,
			\left[
				\begin{array}{c}
					e_1\\
					e_2\\
					\hline
					f^3
				\end{array}
			\right]%
			\label{eq:oscillator_dirac}
			\\[0.25em]
			f^3
			\: &= \:
			d \, e_3
			\,.%
			\label{eq:oscillator_resistive_rel}
		\end{align}%
		\label{eq:oscillator}%
	\end{subequations}
	\Cref{eq:oscillator_storage}
	defines the storage port
	(blue components).
	The matrix in~\cref{eq:oscillator_dirac}
	defines the Dirac structure (green component),
	see~\cref{def:dirac_structure}.
	Its first two rows (and columns) are related to storage
	and thus to the bundle $T \mathcal{X} \oplus T^* \mathcal{X}$.
	The last row is related to
	the trivial bundle
	$\mathbb{R} \times \mathbb{R}^* \rightarrow \mathcal{X}$
	which also appears in
	the definition of the resistive structure (red component)
	based on~\cref{eq:oscillator_resistive_rel},
	see~\cref{def:resistive_structure}.
	Damper and mass interact
	via the power-conjugate variables
	$e_3 = \upsilon$ (velocity)
	and
	$f^3 = d \, \upsilon$ (damping force).
	The Hamiltonian $H \colon \mathcal{X} \rightarrow \mathbb{R}$
	is the same as in~\cref{ex:harmonic_oscillator}.
	Since $H$ is bounded from below,
	the system is passive.%
	\label{ex:oscillator}
\end{example}

The damper represents an irreversible process
that conserves energy as
it turns work into heat.
However, \cref{fig:oscillator_bond_graph} shows the damper
as a one-port component
and the power $e_3 \, f^3$
going into it disappears,
meaning that it is not balanced by
an outgoing power of equal amount.
Thus, there is an obvious discrepancy
between the port-Hamiltonian structure
and the first law of thermodynamics,
at least as long as we believe that
the Hamiltonian represents energy in the thermodynamic sense.

For classical port-Hamiltonian systems
we can argue from a thermodynamics viewpoint as follows:
A reversible interaction
with a thermal reservoir of constant temperature
is (implicitly) assumed
in the modelling process.
The interaction maintains thermal equilibrium
of the system and the (waste heat) reservoir
(i.e.~environment),
thereby making the overall system isothermal.

In~\cite{2009DuindamMacchelliStramigioliBruyninckx} (p.~25),
it is stated that
the physical meaning of the Hamiltonian
(of a classical dissipative port-Hamiltonian system)
is `free energy',
rather than energy.
In equilibrium thermodynamics,
the Helmholtz free energy
is a thermodynamic potential
obtained from the internal energy
via a Legendre transformation with respect to entropy.
A potential contains
all thermodynamic information about
the behaviour of a material at equilibrium,
see e.g.~\cite{2018PavelkaKlikaGrmela} (p.~10).
The thermodynamic potential named after Helmholtz
is called a free energy because
the maximum entropy principle
applied to an isothermal and isochoric non-equilibrium system
implies the minimization of its free energy.
The difference between its free energy in the initial state
and its free energy in the equilibrium state
corresponds to
the maximum (reversible) work production
which can occur as the system passes
from the initial to the equilibrium state
while interacting with
the isothermal reservoir at the \textit{same} temperature,
see e.g.~\cite{1985Callen} (ch.~6).
The statement that the Hamiltonian is a `free energy'
can thus be explained as follows:
An (irrelevant) additive constant in the Hamiltonian
may be identified with
the combined Helmholtz free energy
corresponding to all
(neglected) internal energy storage
of the overall system.
Hence, the term `free energy',
as used in~\cite{2009DuindamMacchelliStramigioliBruyninckx},
additively combines
electro-mechanical energy components
and (constant) Helmholtz free energy components
corresponding to internal energy storage
in the isothermal system and environment.
The electro-mechanical energy components have no entropy content
since all related degrees of freedom are resolved by the model.
Therefore, they are not Legendre-transformed quantities.
In this (perhaps not obvious) sense,
the Hamiltonian is a (Helmholtz) free energy.
The GENERIC literature
also mentions that
Helmholtz free energy
can be used as a single generator
for isothermal systems~%
\cite{2018PavelkaKlikaGrmela} (p.~136).

The idea of summing
different electro-mechanical and `free energy' components
is more commonly understood
within the more general exergy concept.
Indeed, for isothermal and isochoric systems,
the total Helmholtz free energy
essentially coincides with exergy.
Similarly, for isothermal and isobaric systems,
the total Gibbs free energy
essentially coincides with exergy,
see~\cite{1951Keenan}.

Inspired by bond-graph modelling,
the wish to model non-isothermal systems
and to fully express the first law
within the port-Hamiltonian framework
first led to the use of lossless systems,
see e.g.~\cite{2009DuindamMacchelliStramigioliBruyninckx}.
For~\cref{ex:oscillator},
this means that a thermal port is added to the damper,
making it a power-conserving component.
Obviously, the passivity property of being lossless
relates only to the first law of thermodynamics
and ignores the second law of thermodynamics
and all its implications.

\subsection{Isothermal exergetic port-Hamiltonian systems}%

In contrast,
exergetic port-Hamiltonian systems are
coherent with
both the first and the second law of thermodynamics
and
link passivity to degradation of energy.
We consider the oscillator from~\cref{ex:oscillator}
as an exergetic port-Hamiltonian system:

\begin{example}[exergetic model of the damped harmonic oscillator]
	Let us explicitly assume that
	the system is isothermal
	because it is in thermal equilibrium
	with its environment having constant temperature $\theta_0$.
	We consider the environment as
	an (infinitely large) thermal reservoir.
	We need not consider for instance
	its volume, mass or chemical composition
	because at this point only thermal interaction
	with the environment is relevant.
	From the isothermal condition
	\begin{equation*}
		\theta_0 \overset{!}{=} \theta(s) = \pdv{U(s)}{s}
		\,,
	\end{equation*}
	it follows that
	$u =  U(s)  = \theta_0 \, s$
	is a fundamental equation for the environment.
	Its exergy content with respect to itself is
	$U(s) - \theta_0 \, s = 0$,
	see~\cref{eq:exergy_s}.
	Thus, the Hamiltonian already is
	(or at least can be interpreted as)
	an exergetic storage function.

	Since the damper remains at $\theta_0$,
	its exergetic heating power is zero.
	Therefore, it might seem reasonable to omit its thermal port,
	cf.~\cite{2009DuindamMacchelliStramigioliBruyninckx} (p.~25).
	However, for exergetic port-Hamiltonian systems,
	this is made (or kept) explicit,
	as shown in~\cref{fig:oscillator_bond_graph_isothermal}:

	\begin{figure}[!htb]
		\centering
		\includegraphics[width=\textwidth]{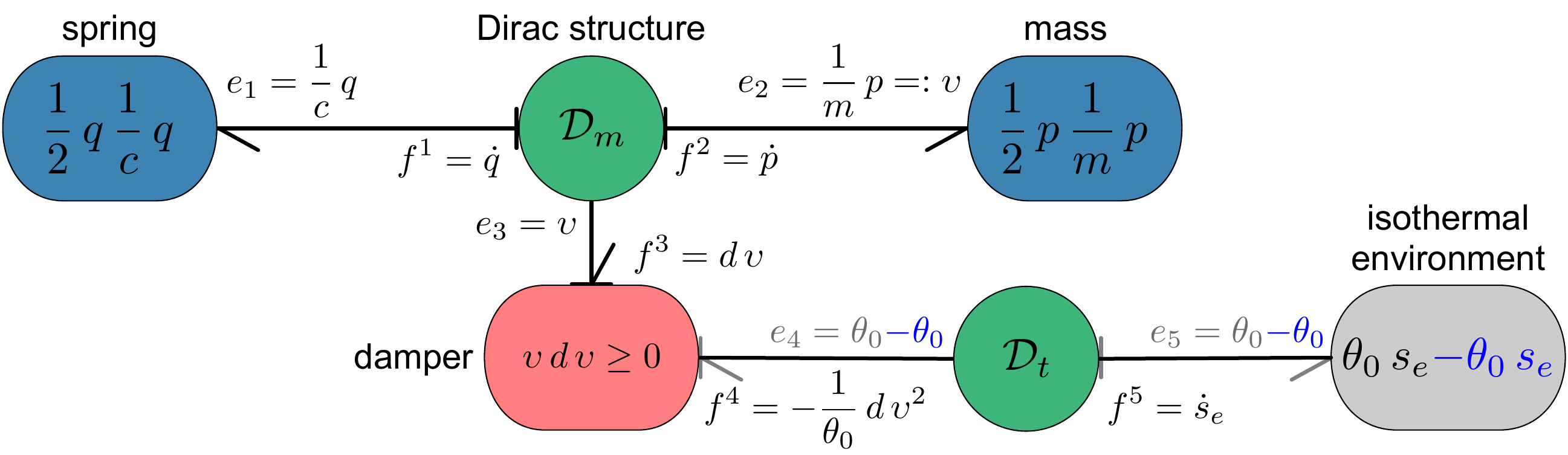}
		\caption{%
			The exergetic port-Hamiltonian model includes
			the reversible exchange of heat
			between the damper and the isothermal environment
			via the Dirac structure $\mathcal{D}_t$.%
		}%
		\label{fig:oscillator_bond_graph_isothermal}
	\end{figure}
	
	The environment is
	like a storage component
	containing zero exergy.
	According to the assumption,
	the damper is held at the environment temperature
	($e_4 = \theta_0 - \theta_0 = 0$).
	Consequently,
	the exergetic power
	$e_4 \,f^4 = e_5 \, f^5$
	vanishes.
	The net power at the damper
	$e_3 \, f^3 + e_4 \, f^4 = d \, \upsilon^2$
	is its exergy destruction rate (or dissipated power).

	The state
	$x = \left( q, \, p, \, s_e \right)
	\in \mathcal{X} = \mathbb{R}^3$
	evolves according to
	\begin{subequations}
	\begin{align}
		\left( f^1, \, f^2, \, f^5, \, e_1, \, e_2, \, e_5 \right)
		\: &= \:
		\left(
			\dot{q}, \,
			\dot{p}, \,
			\dot{s}_e, \,
			\frac{\partial H}{\partial q}(x), \,
			\frac{\partial H}{\partial p}(x), \,
			0
		\right)
		\label{eq:exergetic_oscillator_storage}
		\\[0.25em]
		\left[
			\begin{array}{c}
				f^1\\
				f^2\\
				\hline
				e_3
			\end{array}
		\right]
		\: &= \:
		\left[
			\begin{array}{rr|r}
				 0 & 1 &  0\\
				-1 & 0 & -1\\
				\hline
				 0 & 1 &  0
			\end{array}
		\right]
		\,
		\left[
			\begin{array}{c}
				e_1\\
				e_2\\
				\hline
				f^3
			\end{array}
		\right]%
		\label{eq:exergetic_oscillator_dirac_mechanical}
		\\[0.25em]
		\left[
			\begin{array}{c}
				f^5\\
				\hline
				e_4
			\end{array}
		\right]
		\: &= \:
		\left[
			\begin{array}{r|r}
				 0 & -1\\
				\hline
				 1 &  0
			\end{array}
		\right]
		\,
		\left[
			\begin{array}{c}
				e_5\\
				\hline
				f^4
			\end{array}
		\right]%
		\label{eq:exergetic_oscillator_dirac_thermal}
		\\[0.25em]
		\left[
			\begin{array}{c}
				f^3 \\
				f^4
			\end{array}
		\right]
		\: &= \:
		\frac{1}{\theta_0} \, d \,
		\left[
			\begin{array}{rr}
				 \theta_0   & -\upsilon \\
				-\upsilon & \frac{{\upsilon}^2}{\theta_0}
			\end{array}
		\right]
		\,
		\left[
			\begin{array}{c}
			e_3 \\
			e_4
			\end{array}
		\right]%
		\label{eq:exergetic_oscillator_resistive}
	\end{align}%
	\label{eq:exergetic_oscillator}%
	where $\upsilon = \frac{1}{m} \, p$.
	\end{subequations}
	\Cref{eq:exergetic_oscillator_storage}
	defines the storage port.
	The mass and the spring
	are mechanical components
	containing pure exergy.
	Therefore, no shift appears
	in their contribution to the Hamiltonian.
	The last component of the storage port
	corresponds to the environment.
	\Cref{eq:exergetic_oscillator_dirac_mechanical}
	defines the $\mathcal{D}_m$ component
	($m$ for mechanical)
	and \cref{eq:exergetic_oscillator_dirac_thermal}
	defines the $\mathcal{D}_t$ component
	($t$ for thermal)
	of the Dirac structure.
	Finally,~\cref{eq:exergetic_oscillator_resistive}
	defines the resistive structure.

	While the diagrammatic expression
	for the mathematical model structure
	might seem unnecessarily complicated at first,
	the $\mathcal{D}_t$ component
	represents that the interaction of the damping process
	and the environment is a reversible one.
	At $\mathcal{D}_t$,
	local conservation of entropy holds ($f^5 = -f^4$).

	Regarding the syntax,
	it is not meaningful to directly connect
	a blue component (exposing a storage port)
	with a red component (defining resistive structure).
	The green components
	(defining reversible interconnection)
	mediate all exergy exchange.

	By removing the shifts written in blue
	in~\cref{fig:oscillator_bond_graph_isothermal},
	we obtain a lossless energetic port-Hamiltonian system,
	asserting that the damper conserves energy,
	see~\cref{rem:energetic_bond_graphs}.
	Equivalently, we can assert that
	$\left( e_3, \, e_4 \right)
	= \left( \upsilon, \, \theta_0 \right)$
	lies in the kernel of the matrix
	in~\cref{eq:exergetic_oscillator_resistive}.

	\Cref{eq:exergetic_oscillator} reduces to
	\begin{subequations}
	\begin{align}
		\dot{q} &= \upsilon\\
		\dot{p} &= -\frac{1}{c} \, q - d \, \upsilon\\
		\dot{s}_e &= \frac{1}{\theta_0} \, d \, \upsilon^2
		\,.
	\end{align}%
	\label{eq:exergetic_oscillator_reduced}%
	\end{subequations}
	However, for the seemingly simpler system in~%
	\cref{eq:exergetic_oscillator_reduced},
	checking thermodynamic consistency
	is much harder, especially for a computer.
	For more complex, practical examples,
	a structured and compositional modelling framework
	with a diagrammatic syntax
	is clearly superior.%
	\label{ex:exergetic_oscillator}
\end{example}

We arrive at the following conclusion:
If we assume that
a classical dissipative port-Hamiltonian system
is in thermal equilibrium with
its isothermal (and isobaric) environment
then its Hamiltonian
represents the exergy
of the overall system.
The suggested framework
thus is a straigtforwad extension
of the classical theory.

\begin{remark}[causal strokes]
	Causal strokes do not indicate physical causality
	but mark how information propagates
	when using an explicit time-integration scheme.
	Hence, at storage components, flows must be inputs.
	If a consistent computational causality assignment
	is not possible for all storage components,
	the model yields
	an implicit system of differential-algebraic equations (DAE).
	For instance, placing two capacitors directly in parallel
	results in an algebraic constraint
	demanding equality of voltages.
	In this case, inconsistent initial conditions for the DAE system
	are related to the `two capacitor paradox'.
	The degeneracy could be avoided by taking into account
	the resistance of the wire that connects the capacitors.%
	\label{rem:causal_strokes}
\end{remark}

\begin{remark}[input-output systems]
	The matrix in~\cref{eq:oscillator_dirac}
	is of the form
	\begin{equation*}
		\left[
		\begin{array}{c|c}
			L^\sharp & -g \\
			\hline{}
			g^* & 0
		\end{array}
		\right]
		\,.
	\end{equation*}
	According to~\cref{def:dirac_structure},
	it defines a Dirac structure on
	$T \mathcal{X} \oplus \mathcal{F}_R$
	where $\mathcal{F}_R = \mathcal{X} \times \mathbb{R}$
	is the trivial vector bundle
	on which the resistive structure is defined.
	We focus on the class of input-output systems
	where the bottom-right block is zero.
	The top-left block
	defines a Poisson structure on $\mathcal{X}$
	and thus a Dirac structure
	$\mathcal{D}_{T \mathcal{X}}$
	on $T \mathcal{X}$.
	The top-right block defines
	a vector bundle map
	$-g \colon \mathcal{F}_R \rightarrow T \mathcal{X}$
	covering the identity on $\mathcal{X}$.
	The bottom-left block defines its negative dual $g^*$.
	According to Definition~4.2 in~\cite{2018BarberoCendraGarciaDiego},
	$\left(
		T \mathcal{X}, \, \mathcal{F}_R, \,
		\mathcal{D}_{T \mathcal{X}}, \mathcal{F}_R \oplus \mathcal{F}_R^*, \,
		g
	\right)$
	is an open forward input-output structure.
	Our impression is that~%
	\cite{2018BarberoCendraGarciaDiego}
	introduces a theoretically very appealing
	and possibly also practically useful framework
	for expressing the interconnection of port-Hamiltonian systems.
	However, Dirac structures
	which can directly connect a resistive and a boundary port
	or two different boundary ports (feed-through)
	are quite important in practice.%
	\label{rem:input_output_systems}
\end{remark}

\begin{remark}[classical bond graphs with thermal port]
	Let us assume for a moment
	that the annotated bond-graph expressions
	shown in this work
	are merely figures,
	rather than
	(yet to be defined)
	mathematical objects in their own right.
	Then, we could say that
	removing all shifts
	in the annotated components of the storage function
	and in the annotated expressions for the effort variables
	manipulates such a figure
	into a classical energetic bond graph
	corresponding to a lossless port-Hamiltonian system
	whose passivity property
	is tantamount to the first law of thermodynamics only.
	However, there is no corresponding
	straightforward modification
	of the equations defining the resistive structure.
	This is no surprise
	since lossless systems have no resistive structure.
	The red boxes must turn into
	power-preserving transformers.
	Of course,
	the given annotated expressions
	for the flows at the red boxes
	can straightforwardly be manipulated into
	the equations defining these transformers
	by writing them in terms of
	the efforts (without shifts) at the red boxes.
	It is important to note that
	the resulting equations
	do not follow from
	a structured representation
	enjoying certain properties,
	as it is the case for resistive structures
	in the exergetic port-Hamiltonian framework.%
	\label{rem:energetic_bond_graphs}
\end{remark}

\section{Non-isothermal systems}%
\label{sec:nonisothermal_systems}

We now come to physical modelling of non-isothermal systems.
The following example shows that
the Carnot efficiency naturally appears
in the pairing of
effort and flow variables of a bond representing
(reversible) exchange of heat.

\begin{example}[Carnot engine]
	\begin{figure}[!htb]
		\centering
		\includegraphics[width=\textwidth]{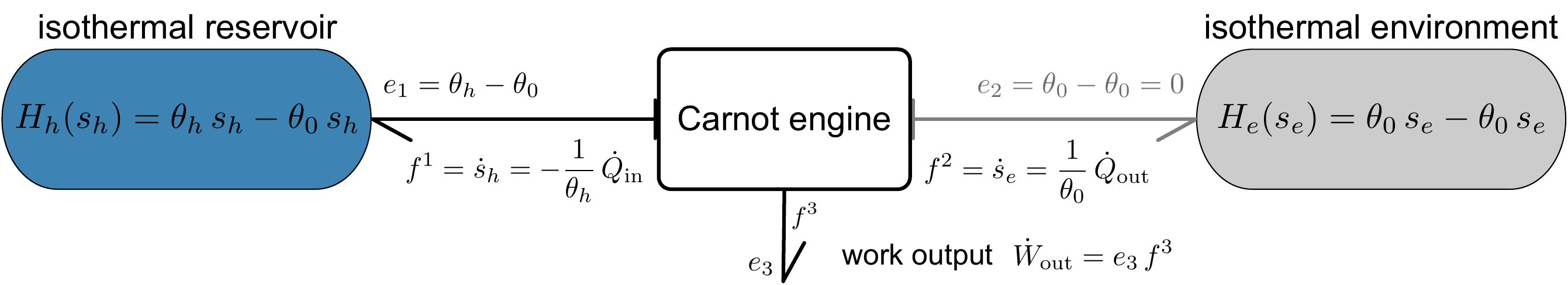}
		\caption{Bond-graph expression of a Carnot engine.}%
		\label{fig:carnot_bond_graph}
	\end{figure}

	\Cref{fig:carnot_bond_graph}
	shows a bond-graph expression of a Carnot engine
	operating between a thermal reservoir at temperature $\theta_h$
	and the environment at temperature $\theta_0$
	which serves as a reference
	to define the exergetic storage function $H = H_h + H_e = H_h$.
	Again, by ignoring the shifts,
	we obtain the energetic power balance
	$\dot{Q}_\text{in} = \dot{Q}_\text{out} + \dot{W}_\text{out}$.
	Due to reversible operation\footnote{%
		There are no resistive components
		hiding inside the abstract white box
		representing the Carnot engine.
	},
	the exergetic power $-e_1 \, f^1$ going into the engine
	is fully converted into work $e_3 \, f^3$:
	\begin{equation*}
		e_3 \, f^3 =
		-e_1 \, f^1 =
		-\pdv{H_h(s_h)}{s_h} \, \dot{s}_h =
		-\left( \theta_h - \theta_0 \right)
		\left( - \frac{1}{\theta_h} \, \dot{Q}_\text{in} \right) =
		\eta_\text{Carnot} \, \dot{Q}_\text{in}
	\end{equation*}

	According to the first law of thermodynamics,
	all physical systems are cyclo-lossless
	if the storage function represents energy.
	In contrast,
	only perfectly reversible devices,
	like the Carnot engine,
	are cyclo-lossless
	if the storage function represents exergy.%
	\label{ex:carnot_bond_graph}
\end{example}

For defining the exergetic storage function
of a thermodynamic system,
we need to know its internal energy as a function of entropy.
In~\cref{ex:carnot_bond_graph}
the thermal reservoir
has a linear and thus unbounded energy function,
reflecting infinite capacity.
Alternatively, we could use entropy as a function of internal energy.
The next example compares the two choices:

\begin{example}[gas-filled compartment]
	\begin{figure}[!htb]
		\centering
		\begin{subfigure}{.5\textwidth}
			\centering
			\includegraphics[height=150pt]{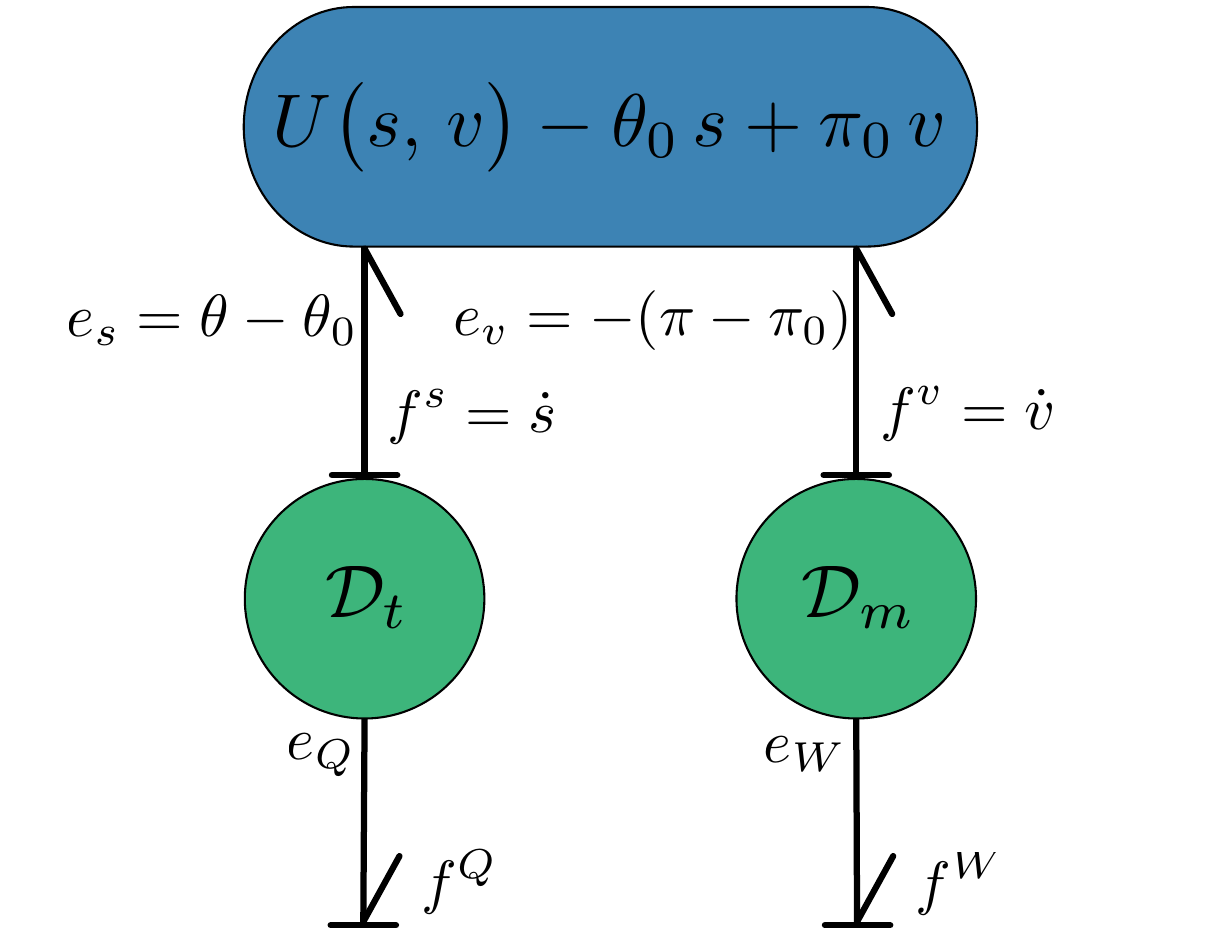}
			\caption{entropy as state variable}%
			\label{fig:compartment_energetic}
		\end{subfigure}%
		\begin{subfigure}{.5\textwidth}
			\centering
			\includegraphics[height=150pt]{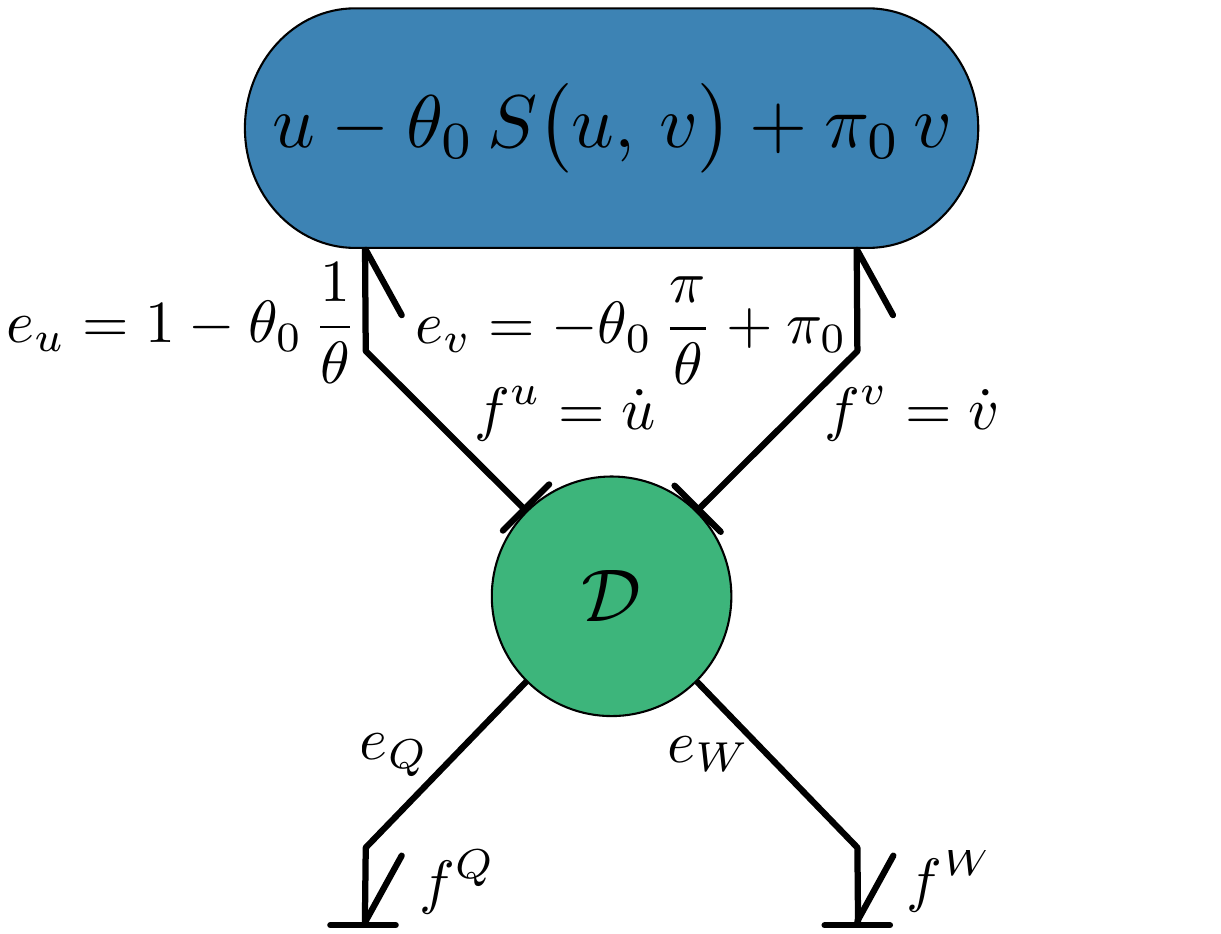}
			\caption{internal energy as state variable}%
			\label{fig:compartment_entropic}
		\end{subfigure}
		\caption{Bond-graph expressions of
			a compartment with heat and work exchange.}%
		\label{fig:compartment}
	\end{figure}

	\Cref{fig:compartment}~shows two bond-graph expressions
	for a gas-filled compartment
	with boundary ports for the exchange of
	thermal exergetic power $e_Q \, f^Q$
	and work $e_W \, f^W$.
	$f^Q$ is the rate of entropy leaving the system,
	$f^W$ is its rate of compression,
	$e_Q$ is its temperature,
	and $e_W$ is its negative pressure,
	both relative to the environment.
	\Cref{fig:compartment_energetic} uses
	entropy and volume as state variables.
	Since exchange of heat/work does not change volume/entropy,
	$\mathcal{D}_t$ and $\mathcal{D}_m$
	have the same trivial form as $\mathcal{D}_t$
	in~\cref{fig:oscillator_bond_graph_isothermal}
	and~\cref{eq:exergetic_oscillator_dirac_thermal}.
	\Cref{fig:compartment_entropic} uses
	internal energy and volume as state variables.
	Exchange of work changes both variables, leading to coupling.
	Since entropy is used as a potential,
	$\mathcal{D}$ is modulated according to
	$\pdv{S}{u} = \frac{1}{\theta}$,
	$\pdv{S}{v} = \frac{\pi}{\theta}$.
	It is defined by
	\begin{equation*}
		\left[
			\begin{array}{c}
				f^u \\
				f^v \\
				\hline
				e_Q \\
				e_W
			\end{array}
		\right]
		\: = \:
		\left[
			\begin{array}{cc|cc}
				0 & 0 &
				-1 / \left( \frac{1}{\theta} \right) &
				\left( \frac{\pi}{\theta} \right) / \left( \frac{1}{\theta} \right) \\
				0 & 0 & 0 & -1 \\
				\hline
				1 / \left( \frac{1}{\theta} \right) & 0 & 0 & 0 \\
				-\left( \frac{\pi}{\theta} \right) / \left( \frac{1}{\theta} \right) &
				1 & 0 & 0
			\end{array}
		\right]
		\,
		\left[
			\begin{array}{c}
				e_u \\
				e_v \\
				\hline
				f^Q \\
				f^W
			\end{array}
		\right]
		\,.
	\end{equation*}
	We will henceforth use internal energy
	as a (local) thermodynamic potential,
	since this yields to simpler models.%
	\label{ex:compartment}
\end{example}

Next, we consider again the damped harmonic oscillator
but this time without assuming thermal equilibrium with the environment.

\begin{example}[non-isothermal damped harmonic oscillator]
	In contrast to~\cref{ex:exergetic_oscillator},
	we model the damper with a thermal capacity
	characterized by an energy function $U$,
	allowing it to heat up.
	Further, we model heat transfer
	characterized by a coefficient $\alpha$,
	allowing the damper to cool down again.
	\Cref{fig:damped_oscillator_bond_graph_nonisothermal}
	shows the model.
	Equations defining the Dirac structure
	according to~\cref{def:dirac_structure}
	and the resistive structure
	according to~\cref{def:resistive_structure}
	are annotated.

	\begin{figure}[!htb]
		\centering
		\includegraphics[width=0.85\textwidth]{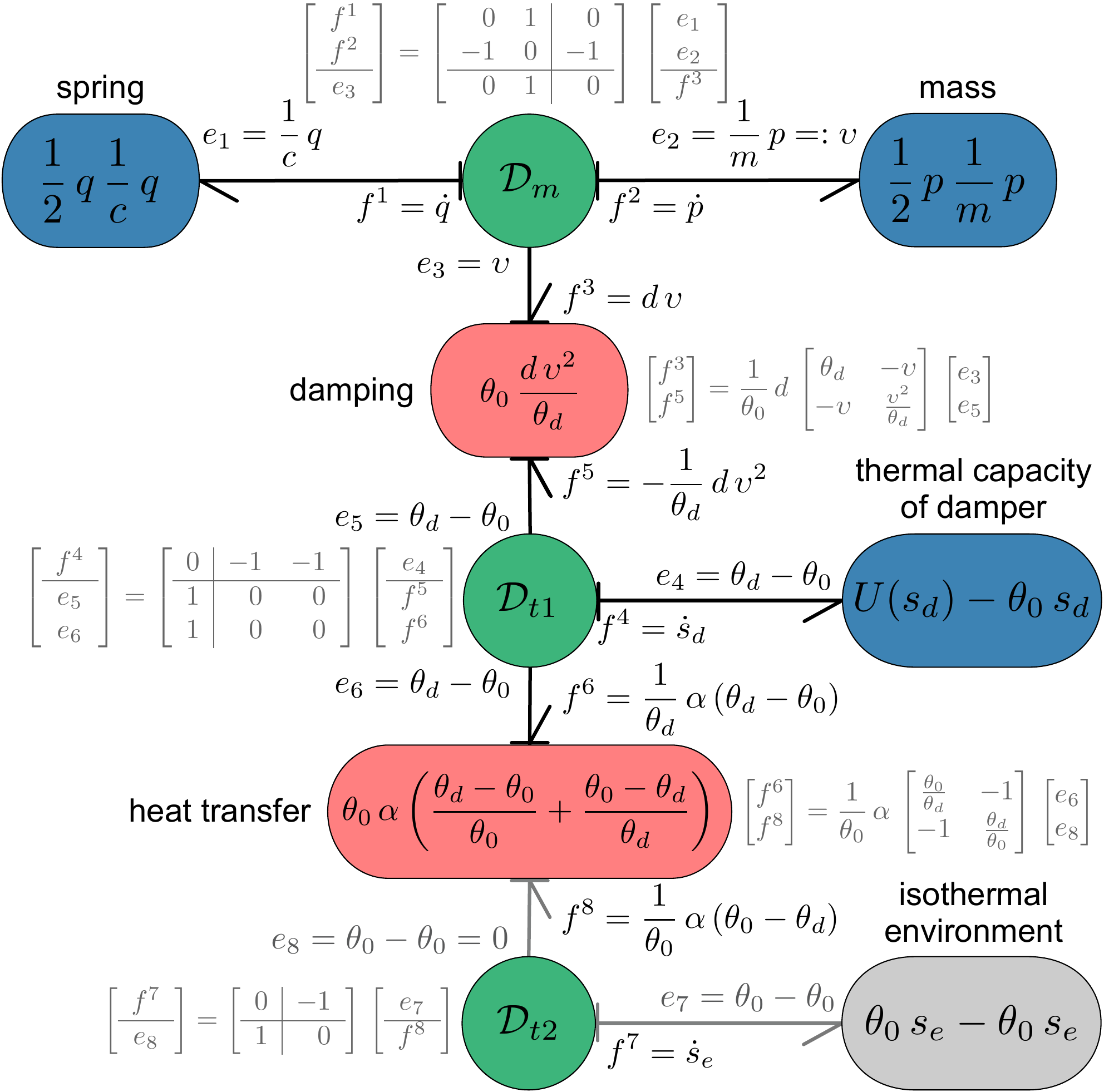}
		\caption{Bond-graph expression of
			a non-isothermal damped harmonic oscillator.}%
		\label{fig:damped_oscillator_bond_graph_nonisothermal}
	\end{figure}

	The damper consumes the mechanical power
	$e_3 \, f^3$
	and produces the thermal exergetic power
	$-e_5 \, f^5 =
	\frac{\theta_d - \theta_0}{\theta_d} \, d \, \upsilon^2$
	which is its heat release rate $d \, \upsilon^2$
	multiplied by the efficiency of a Carnot engine operating between
	the temperature level of
	the damper $\theta_d = \frac{\partial U(s_d)}{\partial s}$
	and that of the environment $\theta_0$.
	The net exergetic power at the damper
	$e_3 \, f^3 + e_5 \, f^5
	= \theta_0 \, \frac{d \, \upsilon^2}{\theta_d}$
	is its exergy destruction rate
	which is its entropy production rate
	$\frac{d \, \upsilon^2}{\theta_d}$
	multiplied by the environment temperature.

	Once $\theta_d > \theta_0$,
	heat is irreversibly transferred
	from the damper's thermal capacity to the environment,
	destroying exergy at the rate
	$e_6 \, f^6 + e_8 \, f^8$.
	We have $e_8 \, f^8 = 0$
	because heat at the environment temperature
	cannot be used to do work.%
	\label{ex:nonisothermal_damped_oscillator}
\end{example}

In summary,
the green components define
the Dirac structure
which encodes reversible (lossless) exchange of exergy.
The net exergetic power at every green component is zero.
Energy and entropy are conserved.
The red components define
the resistive structure
which encodes irreversible processes (relaxation).
At every red component,
the net energetic power is zero.
At the same time, entropy is produced (or conserved),
implying a loss (or conservation) of exergy.
Cyclo-passivity consequently corresponds to the fact that
exergy is either conserved or destroyed.

We now model an isolated cylinder-piston device
both as an exergetic port-Hamiltonian system
and as a GENERIC system.

\begin{example}[isolated cylinder-piston device]
	\begin{figure}[!htb]
		\centering
		\includegraphics[width=0.86\textwidth]{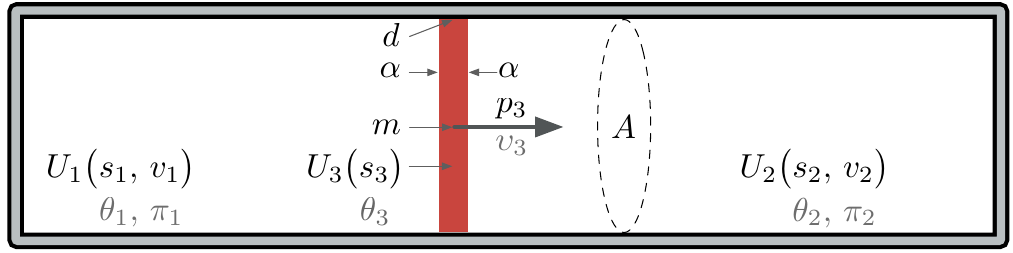}
		\caption{%
			Isolated cylinder
			of cross-sectional area $A$
			containing a piston with mass $m$ and momentum $p_3$.
			The gas in both compartments
			exchanges heat (coefficient $\alpha$)
			with the piston.
			When the piston moves,
			heat is added to it
			due to friction (coefficient $d$).}%
		\label{fig:piston}
	\end{figure}

	\begin{figure}[!htb]
		\centering
		\includegraphics[width=\textwidth]{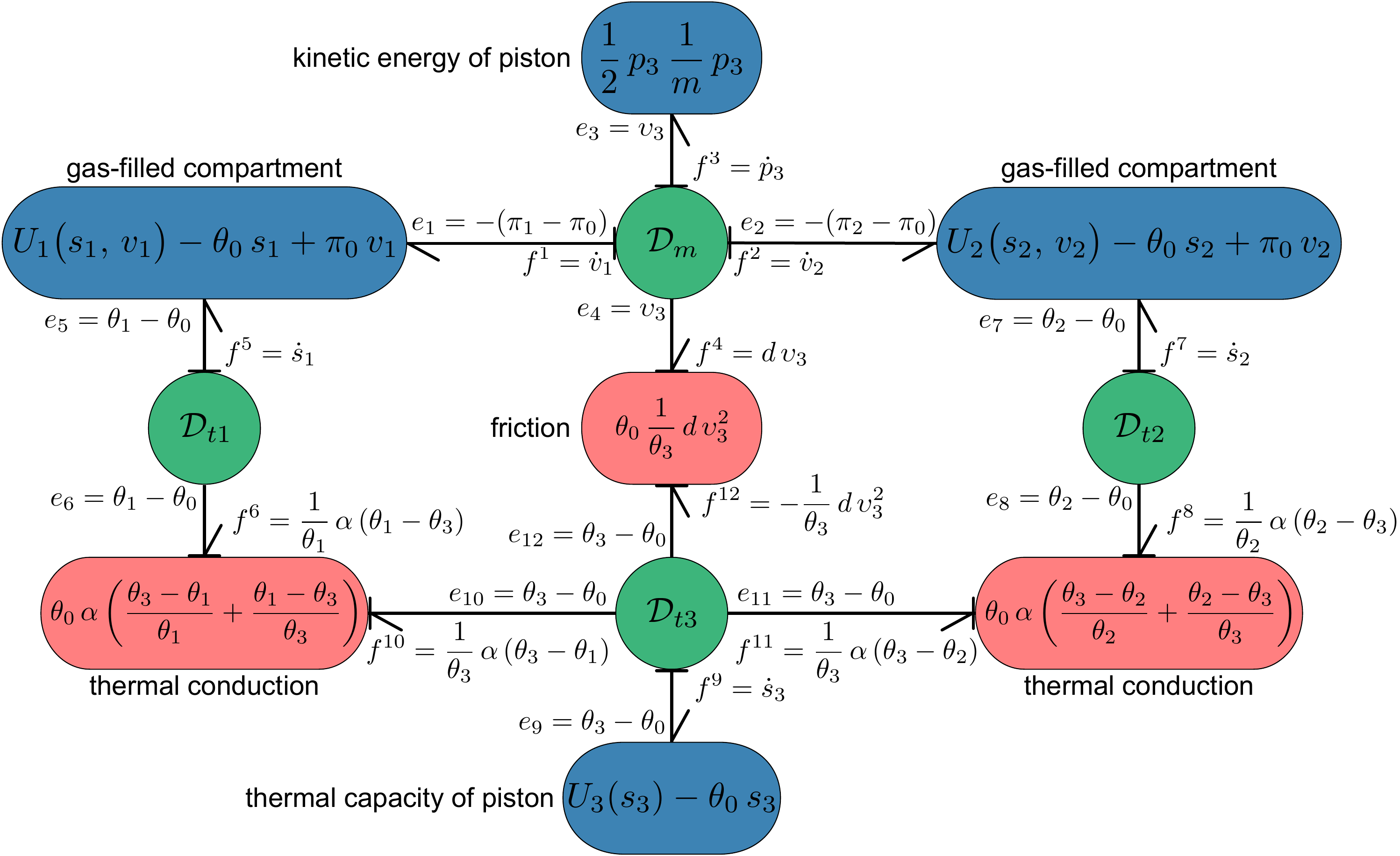}
		\caption{Bond-graph expression of
			the isolated cylinder-piston device in~\cref{fig:piston}.}%
		\label{fig:piston_bond_graph}
	\end{figure}

	\Cref{fig:piston} shows a system comprising of
	two compartments filled with a fixed amount of ideal gas
	which are separated by a piston.
	Its state is
	\begin{equation*}
		x =
		\left( s_1, \, v_1, \, s_2, \, v_2, \, s_3, \, p_3 \right)
		\in \mathcal{X} \subset \mathbb{R}^6
		\,.
	\end{equation*}
	Given that $U_1$, $U_2$, and $U_3$
	are expressions for the internal energy of
	the two compartments\footnote{%
		Functions $U_1$ and $U_2$
		can be derived from the Sackur-Tetrode equation.
	}
	and the piston,
	we define functions $E, \, S, \, V, \, H \in C^\infty(\mathcal{X})$
	for total energy, entropy, volume and exergy by
	\begin{subequations}
		\begin{align}
			E(x) &=
			U_1(s_1, \, v_1) + U_2(s_2, \, v_2) +
			U_3(s_3) + {(p_3)}^2 / (2 \, m) \\
			S(x) &= s_1  + s_2 + s_3 \\
			V(x) &= v_1  + v_2 \\
			H(x) &= E(x) - \theta_0 \, S(x) + \pi_0 \, V(x) + \mathrm{const.}
		\end{align}
	\end{subequations}
	where $\theta_0, \, \pi_0 \in \mathbb{R}^+$
	are the (now arbitrary) reference temperature and pressure,
	see~\cref{eq:exergy_svqp}.

	The Dirac structure $\mathcal{D}_m$ is defined by
	\begin{equation*}
		\left[
			\begin{array}{c}
				f^1 \\
				f^2 \\
				f^3 \\
				\hline
				e_4
			\end{array}
		\right]
		=
		\left[
			\begin{array}{rrr|r}
				 0 & 0 &  A &  0 \\
				 0 & 0 & -A &  0 \\
				-A & A &  0 & -1 \\
				\hline
				 0 & 0 &  1 &  0
			\end{array}
		\right]
		\,
		\left[
			\begin{array}{c}
				e_1 \\
				e_2 \\
				e_3 \\
				\hline
				f^4
			\end{array}
		\right]
		.
	\end{equation*}
	The degeneracy of the top-left block
	corresponds to the (symplectic) Casimir $v_1 + v_2$.
	The Dirac structures
	$\mathcal{D}_{t1}$, $\mathcal{D}_{t2}$ and $\mathcal{D}_{t3}$
	are defined basically as in~\cref{ex:nonisothermal_damped_oscillator}.
	
	The resistive structure corresponding to the damping
	is defined by
	\begin{equation*}
		\begin{bmatrix}
			f^4 \\
			f^{12}
		\end{bmatrix}
		=
		\underbrace{
			\frac{1}{\theta_0} \, d \,
			\begin{bmatrix}
				 \theta_3   & -\upsilon_3 \\
				-\upsilon_3 & \frac{{(\upsilon_3)}^2}{\theta_3}
			\end{bmatrix}
		}_{R(x)}
		\,
		\begin{bmatrix}
			e_4 \\
			e_{12}
		\end{bmatrix}
		=
		\underbrace{
			\begin{bmatrix}
				 \frac{1}{\upsilon_3} \\
				-\frac{1}{\theta_3}
			\end{bmatrix}
		}_{C(x)}
		\underbrace{
			\begin{bmatrix}
				\frac{1}{\theta_0} \, d \,
				{(\upsilon_3)}^2 \, \theta_3
			\end{bmatrix}
		}_{D(x)}
		\underbrace{
			\begin{bmatrix}
				 \frac{1}{\upsilon_3} &
				-\frac{1}{\theta_3}
			\end{bmatrix}
		}_{C^\mathrm{T}(x)}
		\,
		\underbrace{
			\begin{bmatrix}
				\upsilon_3 \\
				\theta_3 - \theta_0
			\end{bmatrix}
		}_{e}
		.
	\end{equation*}
	The factorization
	$R(x) = C(x) \, D(x) \, C^\mathrm{T}(x)$
	shows positive semidefiniteness.
	Energy is conserved since
	$C^\mathrm{T}(x) \,
	\begin{bmatrix} \upsilon_3 & \theta_3 \end{bmatrix}^\mathrm{T} = 0$.
	The remaining term $-\theta_0$ in $e_{12}$ comes from
	$-\theta_0 \, S(x)$ in $H$,
	reflecting that
	the entropy function generates the gradient dynamics.
	The resistive structure corresponding to
	thermal conduction between the left compartment and the piston
	is defined by
	\begin{equation*}
		\begin{bmatrix}
			f^6 \\
			f^{10}
		\end{bmatrix}
		=
		\frac{1}{\theta_0} \, \alpha \,
		\begin{bmatrix}
			\frac{\theta_3}{\theta_1} & -1 \\
			-1 & \frac{\theta_1}{\theta_3}
		\end{bmatrix}
		\,
		\begin{bmatrix}
			e_6 \\
			e_{10}
		\end{bmatrix}
		=
		\begin{bmatrix}
			 \frac{1}{\theta_1} \\
			-\frac{1}{\theta_3}
		\end{bmatrix}
		\begin{bmatrix}
			\frac{1}{\theta_0} \, \alpha \,
			{\left( \sqrt{\theta_1 \, \theta_3} \right)}^2
		\end{bmatrix}
		\begin{bmatrix}
			 \frac{1}{\theta_1} &
			-\frac{1}{\theta_3}
		\end{bmatrix}
		\,
		\begin{bmatrix}
			\theta_1 - \theta_0 \\
			\theta_3 - \theta_0
		\end{bmatrix}
		.
	\end{equation*}
	The factorization has a physical interpretation in terms of LIT:
	$C^\mathrm{T}(x) \, e$ is the thermodynamic force,
	$D(x)$ is the linear relation containing
	the relaxation/transport coefficients,
	and thus $D(x) \, C^\mathrm{T}(x) \, e$ is the thermodynamic flux,
	see~\cite{2005Oettinger}.
	We compare this (without much rigor) to
	a weak-formulation of
	distributed-parameter thermal conduction~\cite{2018BadlyanZimmer}
	which we reformulate in
	the language of exterior calculus\footnote{%
		Due to the separation of
		topological (exterior derivative $\dd$)
		and metric (Hodge star $\star$)
		aspects,
		the continuous and discrete settings
		look more alike.%
	}:
	\begin{equation*}
		\langle \varphi, R(x) \, \psi \rangle =
		\int_{\mathcal{Z}}
		\dd \left( \dfrac{1}{\theta} \, \varphi \right)
		\wedge
		\star \left(
			\frac{1}{\theta_0} \, \kappa \, \theta^2 \,
			\dd \left( \dfrac{1}{\theta} \, \psi \right)
		\right)
	\end{equation*}
	Here,
	$\varphi$ and $\psi$ are test functions.
	The latter is essentially a place holder for
	the effort ($0$-form) $e = \theta - \theta_0$.
	The parameter $\kappa$ is the thermal conductivity.
	For a simple spatial domain $\mathcal{Z}$,
	where $A$ is a cross section
	perpendicular to the direction of thermal conduction
	and $l$ is a length in that direction,
	the Hodge star $\star$ essentially becomes $\frac{A}{l}$
	and $\frac{A}{l} \, \kappa$ indeed corresponds to $\alpha$.
	In the lumped-parameter setting,
	the term $\theta^2$ is replaced by
	the squared geometric mean $\theta_1 \, \theta_3$
	of the two known temperatures.

	The combined\footnote{%
		Since a Dirac structure on
		a vector bundle
		$\mathcal{F} \rightarrow \mathcal{X}$
		is a subbundle of
		$\mathcal{F} \oplus \mathcal{F}^* \rightarrow \mathcal{X}$,
		Dirac structures can be combined
		via the direct sum of bundles.
	} Dirac structure
	$\mathcal{D} \cong
	\mathcal{D}_{m} \oplus
	\mathcal{D}_{t1} \oplus
	\mathcal{D}_{t2} \oplus
	\mathcal{D}_{t3}$
	is defined by
	\begin{equation*}
		\left[
			\begin{array}{c}
				f^1 \\
				f^2 \\
				f^3 \\
				f^5 \\
				f^7 \\
				f^9 \\
				\hline
				e_4 \\
				e_{12} \\
				e_6 \\
				e_{10} \\
				e_8 \\
				e_{11}
			\end{array}
		\right]
		=
		\left[
			\begin{array}{rrrrrr|rrrrrr}
				   &   &  A &   &   &   &    &    &    &    &    &    \\
				   &   & -A &   &   &   &    &    &    &    &    &    \\
				-A & A &    &   &   &   & -1 &    &    &    &    &    \\
				   &   &    &   &   &   &    &    & -1 &    &    &    \\
				   &   &    &   &   &   &    &    &    &    & -1 &    \\
				   &   &    &   &   &   &    & -1 &    & -1 &    & -1 \\
				\hline
				   &   &  1 &   &   &   &    &    &    &    &    &    \\
				   &   &    &   &   & 1 &    &    &    &    &    &    \\
				   &   &    & 1 &   &   &    &    &    &    &    &    \\
				   &   &    &   &   & 1 &    &    &    &    &    &    \\
				   &   &    &   & 1 &   &    &    &    &    &    &    \\
				   &   &    &   &   & 1 &    &    &    &    &    &    \\
			\end{array}
		\right]
		\:
		\left[
			\begin{array}{c}
				e_1 \\
				e_2 \\
				e_3 \\
				e_5 \\
				e_7 \\
				e_9 \\
				\hline
				f^4 \\
				f^{12} \\
				f^6 \\
				f^{10} \\
				f^8 \\
				f^{11}
			\end{array}
		\right].
	\end{equation*}
	The top-left block $L$
	defines a Poisson structure $\pb{\cdot}{\cdot}$ on $\mathcal{X}$.
	The top-right block $-g$
	defines a vector bundle map
	$g \colon \mathcal{F}_R \rightarrow T \mathcal{X}$,
	see~\cref{rem:input_output_systems}.
	The block-diagonal combination of
	the three tensors defining the component resistive structures
	defines the combined resistive structure $\mathcal{R}$ on $\mathcal{F}_R$.
	Then
	$\left( \mathcal{X}, \, \mathcal{F}_R, \, \{ 0 \},
	\mathcal{D}, \, \mathcal{R}, \, H \right)$
	is a port-Hamiltonian system.
	Its state evolves according to
	$\dot{x} = \left( L - g \, R(x) \, g^* \right) \dd H$.
	Let $M = \theta_0 \, g \, R \, g^*$
	define a gradient structure $\Psi$ on $\mathcal{X}$.
	Then
	$\left( \mathcal{X}, \, \pb{\cdot}{\cdot}, \, \Psi, \, E, \, S \right)$
	defines a GENERIC system.
	Its state evolves according to
	$\dot{x} = L \, \dd E + M(x) \, \dd S$.%
	\label{ex:piston}
\end{example}

The above shows that
an isolated exergetic port-Hamiltonian system
is equivalent to a GENERIC system
and thus it is thermodynamically consistent.
It is important to note that the systems in~%
\cref{ex:exergetic_oscillator}
or \cref{ex:nonisothermal_damped_oscillator}
are also isolated systems (including their environment).
In the future,
a rigorous compositional framework
will enable us to
assert thermodynamic consistency
for open systems and their interconnection.

In the final example,
we consider
an electrically-heated cylinder-piston device
which exchanges heat and work
with the (isothermal and isobaric) environment.

\begin{example}[heated cylinder-piston device]
	\begin{figure}[!htb]
		\centering
		\includegraphics[width=0.75\textwidth]{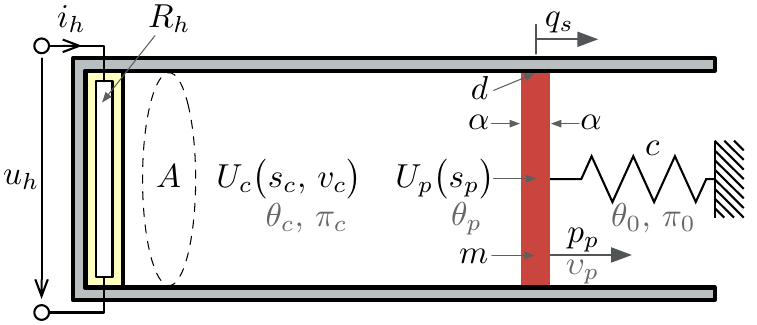}
		\caption{An open cylinder
			with cross-sectional area $A$
			and isolating walls
			contains an electric heater with resistance $R_h$.
			A piston with mass $m$ and momentum $p_p$
			closes the cylinder.
			Attached to it is
			a spring with compliance $c$ and extension $q_s$.
			The gas in the compartment and the environment
			both exchange
			heat (coefficient $\alpha$)
			with the piston.
			When the piston moves,
			heat is added to it
			due to friction (coefficient $d$).}%
		\label{fig:piston2}
	\end{figure}

	\begin{figure}[!htb]
		\centering
		\includegraphics[width=\textwidth]{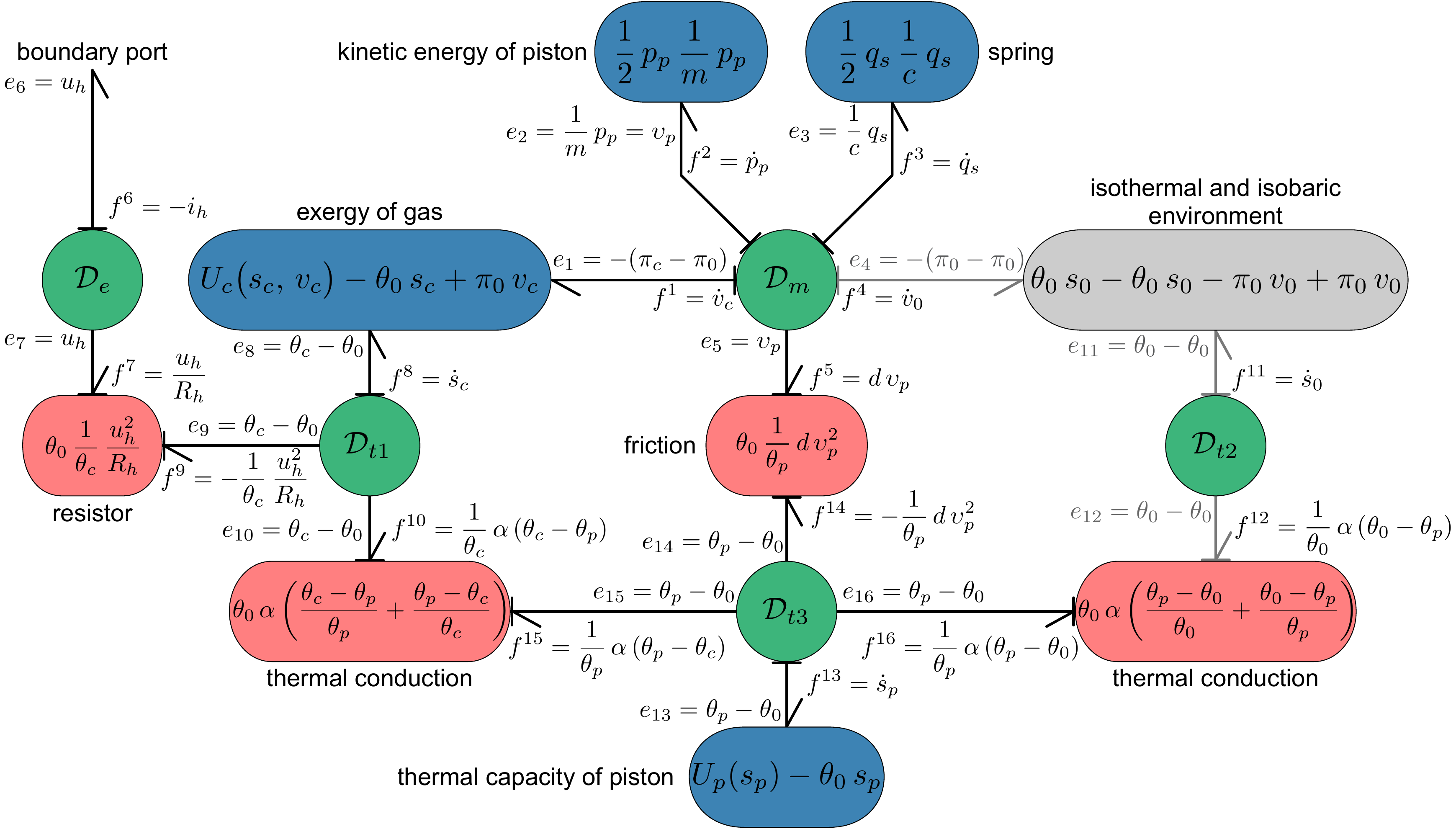}
		\caption{Bond-graph expression of
			the open cylinder-piston device in~\cref{fig:piston2}.}%
		\label{fig:piston2_bond_graph}
	\end{figure}

	\Cref{fig:piston2} depicts the device,
	while
	the bond-graph expression in~\cref{fig:piston2_bond_graph}
	conveys the structure of the corresponding physical model.
	The system is open.
	To determine a dynamics,
	we have to interconnect
	the electric boundary port $\left( f^6, \, e_6 \right)$
	with another system (having the same environment)
	such that the composite forms an isolated system.
	We merely state some details
	which differ from previous examples.

	The Dirac structure $\mathcal{D}_m$ is defined by
	\begin{equation*}
		\left[
			\begin{array}{c}
				f^1 \\
				f^2 \\
				f^3 \\
				f^4 \\
				\hline
				e_5
			\end{array}
		\right]
		=
		\left[
			\begin{array}{rrrr|r}
				 0 &  A &  0 & 0 &  0 \\
				-A &  0 & -1 & A & -1 \\
				 0 &  1 &  0 & 0 &  0 \\
				 0 & -A &  0 & 0 &  0 \\
				\hline
				 0 &  1 &  0 & 0 &  0
			\end{array}
		\right]
		\,
		\left[
			\begin{array}{c}
				e_1 \\
				e_2 \\
				e_3 \\
				e_4 \\
				\hline
				f^5
			\end{array}
		\right]
		.
	\end{equation*}
	The degeneracy of the top-left block
	corresponds to (symplectic) Casimirs
	$v_c + v_0$ and $v_c / A - q_s$.
	When the gas expands,
	it has to displace the isobaric environment.
	The corresponding mechanical work
	cannot be extracted
	and hence $e_4 \, f^4 = 0$.
	Only the exergetic power $e_1 \, f^1$
	could possibly be extracted.

	The resistive structure corresponding to the resistor
	is defined by
	\begin{equation*}
		\begin{bmatrix}
			f^7 \\
			f^9
		\end{bmatrix}
		=
		\frac{1}{\theta_0} \, \frac{1}{R_h} \,
		\begin{bmatrix}
			\theta_c & -u_h \\
			-u_h     & \frac{{(u_h)}^2}{\theta_c}
		\end{bmatrix}
		\,
		\begin{bmatrix}
			e_7 \\
			e_9
		\end{bmatrix}
		=
		\begin{bmatrix}
			 \frac{1}{u_h} \\
			-\frac{1}{\theta_c}
		\end{bmatrix}
		\begin{bmatrix}
			\frac{1}{\theta_0} \, \frac{1}{R_h} \,
			{(u_h)}^2 \, \theta_c
		\end{bmatrix}
		\begin{bmatrix}
			 \frac{1}{u_h} &
			-\frac{1}{\theta_c}
		\end{bmatrix}
		\,
		\begin{bmatrix}
			u_h \\
			\theta_c - \theta_0
		\end{bmatrix}
	\end{equation*}
	which is analogous to the case of friction.%
	\label{ex:piston2}
\end{example}

\section{Conclusions}%
\label{sec:conclusions}

The compositional nature of port-Hamiltonian systems
makes them attractive for
modelling of interconnected and controlled physical systems.
The present framework enhances
classical port-Hamiltonian theory
with
a physically sound interpretation of dissipativity
and a refined structure that ensures thermodynamic consistency.
Future work must address the question
how the interconnection of exergetic port-Hamiltonian systems
can be formalized and implemented
such that thermodynamic consistency
of constituent systems
implies the consistency of composite systems.

At the heart of this work
lies the goal to develop
a practical framework to support
the design and operation of sustainable energy systems.
Not only as a consequence
of increasing demand for sustainable technology,
teams are becoming more interdisciplinary.
This is one important reason why
intuitive abstractions with computational meaning
are of central importance for the future of engineering.
Exergetic port-Hamiltonian systems
admit a diagrammatic syntax
that is derived form bond-graph modelling.
Future work must thus address the question
how expressions in this syntax
can be formalized as mathematical objects
based on which computations
can be performed.
Once this foundation is established,
exergetic port-Hamiltonian systems
can become a valuable tool for
thermodynamic design and optimization.
Their diagrammatic syntax will help humans
to think and communicate,
while their underlying mathematical framework,
together with modern compiler technology,
will efficiently handle
the tedious parts of computational procedures
such as
modular and hierarchical composition,
model transformations,
simulation,
optimization,
and control design.
Since many and eventually possibly all
such procedures will provably preserve
the compositional and the thermodynamic structure,
a lot of effort previously spent on
arranging parts, bookkeeping and verification
can then be spent on sustainable design.

\section*{Acknowledgements}%

Markus Lohmayer wishes to thank
Candan Güdücü,
Hannes Dänschel,
Michal Pavelka,
Riccardo Morandin,
Rodrigo Sato Mart\'in de Almagro,
and Volker Mehrmann
for valuable discussions.

\section*{Disclosure statement}%

No potential conflict of interest was reported by the authors.


\phantomsection
\addcontentsline{toc}{chapter}{References}

\bibliographystyle{link-elsarticle-num}

\bibliography{literature.bib}

\begin{thebibliography}{10}
\expandafter\ifx\csname url\endcsname\relax
  \def\url#1{\texttt{#1}}\fi
\expandafter\ifx\csname urlprefix\endcsname\relax\def\urlprefix{URL }\fi
\expandafter\ifx\csname href\endcsname\relax
  \def\href#1#2{#2} \def\path#1{#1}\fi
\providecommand*{\nolinkurl}{\url}

\bibitem{2005Szargut}
J.~Szargut, Exergy Method: Technical and Ecological Applications, WIT Press,
  2005.

\bibitem{1956Rant}
\href{https://doi.org/10.1007/BF02592661}{Z.~Rant, Exergie, ein neues {W}ort
  f\"ur ``technische {A}rbeitsf\"ahigkeit'', Forschung auf dem Gebiet des
  Ingenieurwesens A 22~(1) (1956) 33--37.
\newblock \path{doi:10.1007/BF02592661}.}

\bibitem{1873Gibbs}
J.~W. Gibbs, A method of geometrical representation of the thermodynamic
  properties of substances by means of surfaces, Transactions of the
  Connecticut Academy of Arts and Sciences 2 (1873) 382--404.

\bibitem{2002GaggioliRichardsonBowman}
\href{https://doi.org/10.1115/1.1448336}{R.~A. Gaggioli, D.~H. Richardson,
  A.~J. Bowman, Available energy{\textemdash}{P}art {I}: Gibbs revisited,
  Journal of Energy Resources Technology 124~(2) (2002) 105--109.
\newblock \path{doi:10.1115/1.1448336}.}

\bibitem{2002GaggioliPaulus}
\href{https://doi.org/10.1115/1.1448337}{R.~A. Gaggioli, D.~M. Paulus,
  Available energy{\textemdash}{P}art {II}: Gibbs extended, Journal of Energy
  Resources Technology 124~(2) (2002) 110--115.
\newblock \path{doi:10.1115/1.1448337}.}

\bibitem{1951Keenan}
\href{https://doi.org/10.1088/0508-3443/2/7/302}{J.~H. Keenan, Availability and
  irreversibility in thermodynamics, British Journal of Applied Physics 2~(7)
  (1951) 183--192.
\newblock \path{doi:10.1088/0508-3443/2/7/302}.}

\bibitem{1980Gaggioli}
\href{https://doi.org/10.1021/bk-1980-0122}{R.~A. Gaggioli (Ed.),
  Thermodynamics: Second Law Analysis, Vol. 122, American Chemical Society,
  1980.
\newblock \path{doi:10.1021/bk-1980-0122}.}

\bibitem{1985Kotas}
\href{https://doi.org/10.1016/c2013-0-00894-8}{T.~Kotas, The Exergy Method of
  Thermal Plant Analysis, Elsevier, 1985.
\newblock \path{doi:10.1016/c2013-0-00894-8}.}

\bibitem{1996Bejan}
\href{https://doi.org/10.1063/1.362674}{A.~Bejan, Entropy generation
  minimization: The new thermodynamics of finite-size devices and finite-time
  processes, Journal of Applied Physics 79~(3) (1996) 1191--1218.
\newblock \path{doi:10.1063/1.362674}.}

\bibitem{2011Andresen}
\href{https://doi.org/10.1002/anie.201001411}{B.~Andresen, Current trends in
  finite-time thermodynamics, Angewandte Chemie International Edition 50~(12)
  (2011) 2690--2704.
\newblock \path{doi:10.1002/anie.201001411}.}

\bibitem{1979aRubin}
\href{https://doi.org/10.1103/physreva.19.1272}{M.~H. Rubin, Optimal
  configuration of a class of irreversible heat engines. {I}, Physical Review A
  19~(3) (1979) 1272--1276.
\newblock \path{doi:10.1103/physreva.19.1272}.}

\bibitem{1979bRubin}
\href{https://doi.org/10.1103/physreva.19.1277}{M.~H. Rubin, Optimal
  configuration of a class of irreversible heat engines. {II}, Physical Review
  A 19~(3) (1979) 1277--1289.
\newblock \path{doi:10.1103/physreva.19.1277}.}

\bibitem{2001SalamonNultonSiragusaAndersenLimon}
\href{https://doi.org/10.1016/S0360-5442(00)00059-1}{P.~Salamon, J.~Nulton,
  G.~Siragusa, T.~Andersen, A.~Limon, Principles of control thermodynamics,
  Energy 26~(3) (2001) 307 -- 319.
\newblock \path{doi:10.1016/S0360-5442(00)00059-1}.}

\bibitem{1992GovernToole1}
J.~A. McGovern, F.~O'Toole, A virtual-system concept for exergy analysis of
  flow network plant; part {I}: Principles, in: Proceedings of International
  Symposium on Efficiency, Costs, Optimization and Simulation of Energy
  Systems, New York, NY, 1992, pp. 155--160.

\bibitem{1990Courant}
\href{https://doi.org/10.1090/S0002-9947-1990-0998124-1}{T.~J. Courant, {D}irac
  manifolds, Transactions of the American Mathematical Society 319~(2) (1990)
  631--661.
\newblock \path{doi:10.1090/S0002-9947-1990-0998124-1}.}

\bibitem{1993Dorfman}
I.~Dorfman, {D}irac Structures and Integrability of Nonlinear Evolution
  Equations, John Wiley, Chichester, 1993.

\bibitem{1998DalsmoSchaft}
\href{https://doi.org/10.1137/s0363012996312039}{M.~Dalsmo, A.~van~der Schaft,
  On representations and integrability of mathematical structures in
  energy-conserving physical systems, {SIAM} Journal on Control and
  Optimization 37~(1) (1998) 54--91.
\newblock \path{doi:10.1137/s0363012996312039}.}

\bibitem{2009Merker}
\href{https://doi.org/10.1007/s00332-009-9052-3}{J.~Merker, On the geometric
  structure of {H}amiltonian systems with ports, Journal of Nonlinear Science
  19~(6) (2009) 717--738.
\newblock \path{doi:10.1007/s00332-009-9052-3}.}

\bibitem{2018BarberoCendraGarciaDiego}
\href{https://doi.org/10.1088/1751-8121/aad4ba}{M.~Barbero-Li{\~{n}}{\'{a}}n,
  H.~Cendra, E.~G.-T. Andr{\'{e}}s, D.~M. de~Diego, New insights in the
  geometry and interconnection of port-{H}amiltonian systems, Journal of
  Physics A: Mathematical and Theoretical 51~(37) (2018) 375201.
\newblock \path{doi:10.1088/1751-8121/aad4ba}.}

\bibitem{1972Willems}
\href{https://doi.org/10.1007/bf00276493}{J.~C. Willems, Dissipative dynamical
  systems part {I}: General theory, Archive for Rational Mechanics and Analysis
  45~(5) (1972) 321--351.
\newblock \path{doi:10.1007/bf00276493}.}

\bibitem{2017Schaft}
\href{https://doi.org/10.1007/978-3-319-49992-5}{A.~van~der Schaft, {L2}-Gain
  and Passivity Techniques in Nonlinear Control, 3rd Edition, Springer
  International Publishing, 2017.
\newblock \path{doi:10.1007/978-3-319-49992-5}.}

\bibitem{2020Schaft}
\href{https://doi.org/10.1109/TAC.2020.3013941}{A.~{van der Schaft},
  Cyclo-dissipativity revisited, IEEE Transactions on Automatic Control (2020).
\newblock \path{doi:10.1109/TAC.2020.3013941}.}

\bibitem{2009DuindamMacchelliStramigioliBruyninckx}
\href{https://doi.org/10.1007/978-3-642-03196-0}{V.~Duindam, A.~Macchelli,
  S.~Stramigioli, H.~Bruyninckx (Eds.), Modeling and Control of Complex
  Physical Systems, Springer Berlin Heidelberg, 2009.
\newblock \path{doi:10.1007/978-3-642-03196-0}.}

\bibitem{2004EberardMaschke}
\href{https://doi.org/10.1016/S1474-6670(17)31230-2}{D.~Eberard, B.~Maschke,
  Port {H}amiltonian systems extended to irreversible systems: The example of
  the heat conduction, IFAC Proceedings Volumes 37~(13) (2004) 243--248.
\newblock \path{doi:10.1016/S1474-6670(17)31230-2}.}

\bibitem{2013RamirezMaschkeSbarbaro}
\href{https://doi.org/10.1016/j.ces.2012.12.002}{H.~Ramirez, B.~Maschke,
  D.~Sbarbaro, Irreversible port-{H}amiltonian systems: a general formulation
  of irreversible processes with application to the {CSTR}, Chemical
  Engineering Science 89 (2013) 223--234.
\newblock \path{doi:10.1016/j.ces.2012.12.002}.}

\bibitem{1973Hermann}
R.~Hermann, Geometry, physics, and systems, M. Dekker, New York, 1973.

\bibitem{2007EberardMaschkeSchaft}
\href{https://doi.org/10.1016/s0034-4877(07)00024-9}{D.~Eberard, B.~Maschke,
  A.~van~der Schaft, An extension of {H}amiltonian systems to the thermodynamic
  phase space: Towards a geometry of nonreversible processes, Reports on
  Mathematical Physics 60~(2) (2007) 175--198.
\newblock \path{doi:10.1016/s0034-4877(07)00024-9}.}

\bibitem{2018SchaftMaschke}
\href{https://doi.org/10.3390/e20120925}{A.~van~der Schaft, B.~Maschke,
  Geometry of thermodynamic processes, Entropy 20~(12) (2018) 925.
\newblock \path{doi:10.3390/e20120925}.}

\bibitem{1980DzyaloshinskiiVolovick}
\href{https://doi.org/10.1016/0003-4916(80)90119-0}{I.~Dzyaloshinskii,
  G.~Volovick, Poisson brackets in condensed matter physics, Annals of Physics
  125~(1) (1980) 67--97.
\newblock \path{doi:10.1016/0003-4916(80)90119-0}.}

\bibitem{1984Grmela}
\href{https://doi.org/10.1090/conm/028/751978}{M.~Grmela, Particle and bracket
  formulations of kinetic equations, in: Fluids and plasmas: geometry and
  dynamics ({B}oulder, {C}olo., 1983), Vol.~28 of Contemp. Math., Amer. Math.
  Soc., Providence, RI, 1984, pp. 125--132.
\newblock \path{doi:10.1090/conm/028/751978}.}

\bibitem{1984Kaufman}
\href{https://doi.org/10.1016/0375-9601(84)90634-0}{A.~N. Kaufman, Dissipative
  {H}amiltonian systems: A unifying principle, Physics Letters A 100~(8) (1984)
  419--422.
\newblock \path{doi:10.1016/0375-9601(84)90634-0}.}

\bibitem{1984Morrison}
\href{https://doi.org/10.1016/0375-9601(84)90635-2}{P.~J. Morrison, Bracket
  formulation for irreversible classical fields, Physics Letters A 100~(8)
  (1984) 423--427.
\newblock \path{doi:10.1016/0375-9601(84)90635-2}.}

\bibitem{1997GrmelaOettinger}
\href{https://doi.org/10.1103/PhysRevE.56.6620}{M.~Grmela, H.~C. Öttinger,
  Dynamics and thermodynamics of complex fluids. {I}. development of a general
  formalism, Phys. Rev. E 56 (1997) 6620--6632.
\newblock \path{doi:10.1103/PhysRevE.56.6620}.}

\bibitem{1997OettingerGrmela}
\href{https://doi.org/10.1103/PhysRevE.56.6633}{H.~C. Öttinger, M.~Grmela,
  Dynamics and thermodynamics of complex fluids. {II.} illustrations of a
  general formalism, Phys. Rev. E 56 (1997) 6633--6655.
\newblock \path{doi:10.1103/PhysRevE.56.6633}.}

\bibitem{2005Oettinger}
H.~C. Öttinger, Beyond Equilibrium Thermodynamics, John Wiley \& Sons Inc,
  2005.

\bibitem{2018PavelkaKlikaGrmela}
\href{https://doi.org/10.1515/9783110350951}{M.~Pavelka, V.~Klika, M.~Grmela,
  Multiscale Thermo-Dynamics, De Gruyter, 2018.
\newblock \path{doi:10.1515/9783110350951}.}

\bibitem{2014PavelkaKlikaGrmela}
\href{https://doi.org/10.1103/PhysRevE.90.062131}{M.~Pavelka, V.~Klika,
  M.~Grmela, Time reversal in nonequilibrium thermodynamics, Phys. Rev. E 90
  (2014) 062131.
\newblock \path{doi:10.1103/PhysRevE.90.062131}.}

\bibitem{1957Jaynes}
\href{https://doi.org/10.1103/PhysRev.106.620}{E.~T. Jaynes, Information theory
  and statistical mechanics, Phys. Rev. 106 (1957) 620--630.
\newblock \path{doi:10.1103/PhysRev.106.620}.}

\bibitem{2019KlikaPavelkaVagnerGrmela}
\href{https://doi.org/10.3390/e21070715}{V.~Klika, M.~Pavelka, P.~Vágner,
  M.~Grmela, Dynamic maximum entropy reduction, Entropy 21~(7) (2019).
\newblock \path{doi:10.3390/e21070715}.}

\bibitem{2018Grmela}
\href{https://doi.org/10.1088/2399-6528/aab642}{M.~Grmela, {GENERIC} guide to
  the multiscale dynamics and thermodynamics, Journal of Physics Communications
  2~(3) (2018) 032001.
\newblock \path{doi:10.1088/2399-6528/aab642}.}

\bibitem{2016MielkeRengerPeletier}
\href{https://doi.org/10.1515/jnet-2015-0073}{A.~Mielke, D.~R.~M. Renger, M.~A.
  Peletier, A generalization of {O}nsager's reciprocity relations to gradient
  flows with nonlinear mobility, Journal of Non-Equilibrium Thermodynamics
  41~(2) (2016).
\newblock \path{doi:10.1515/jnet-2015-0073}.}

\bibitem{1984GrootMazur}
S.~de~Groot, P.~Mazur, Non-equilibrium Thermodynamics, Dover Books on Physics,
  Dover Publications, 1984.

\bibitem{2013HuettnerSvendsen}
\href{https://doi.org/10.1007/s00161-012-0289-y}{M.~Hütter, B.~Svendsen,
  Quasi-linear versus potential-based formulations of force{\textendash}flux
  relations and the {GENERIC} for irreversible processes: comparisons and
  examples, Continuum Mechanics and Thermodynamics 25~(6) (2013) 803--816.
\newblock \path{doi:10.1007/s00161-012-0289-y}.}

\bibitem{2020ShangOettinger}
\href{https://doi.org/10.1098/rspa.2019.0446}{X.~Shang, H.~C. Öttinger,
  Structure-preserving integrators for dissipative systems based on
  reversible{\textendash}irreversible splitting, Proceedings of the Royal
  Society A: Mathematical, Physical and Engineering Sciences 476~(2234) (2020)
  20190446.
\newblock \path{doi:10.1098/rspa.2019.0446}.}

\bibitem{2006Oettinger}
\href{https://doi.org/10.1103/PhysRevE.73.036126}{H.~C. Öttinger,
  Nonequilibrium thermodynamics for open systems, Phys. Rev. E 73 (2006)
  036126.
\newblock \path{doi:10.1103/PhysRevE.73.036126}.}

\bibitem{2018BadlyanZimmer}
A.~M. Badlyan, C.~Zimmer, Operator-{GENERIC} {F}ormulation of {T}hermodynamics
  of {I}rreversible {P}rocesses (2018).
\newblock \path{arXiv:1807.09822}.

\bibitem{1997YdstieAlonso}
\href{https://doi.org/10.1016/S0167-6911(97)00023-6}{B.~E. Ydstie, A.~A.
  Alonso, Process systems and passivity via the {Clausius-Planck} inequality,
  Systems \& Control Letters 30~(5) (1997) 253 -- 264.
\newblock \path{doi:10.1016/S0167-6911(97)00023-6}.}

\bibitem{2011HoangCouenneJallutGorrec}
\href{https://doi.org/10.1016/j.jprocont.2011.06.014}{H.~Hoang, F.~Couenne,
  C.~Jallut, Y.~L. Gorrec, The port {H}amiltonian approach to modeling and
  control of {C}ontinuous {S}tirred {T}ank {R}eactors, Journal of Process
  Control 21~(10) (2011) 1449--1458.
\newblock \path{doi:10.1016/j.jprocont.2011.06.014}.}

\bibitem{2018ZitteHamrounCouennePitault}
\href{https://doi.org/10.1016/j.ifacol.2018.06.012}{B.~Zitte, B.~Hamroun,
  F.~Couenne, I.~Pitault, Representation of heat exchanger networks using graph
  formalism, in: Proceedings of the 6th IFAC Workshop on Lagrangian and
  Hamiltonian Methods for Nonlinear Control (LHMNC), Valpara\'{i}so, Chile,
  2018, pp. 44--49.
\newblock \path{doi:10.1016/j.ifacol.2018.06.012}.}

\bibitem{2018BadlyanMaschkeBeattieMehrmann}
A.~M. Badlyan, B.~Maschke, C.~Beattie, V.~Mehrmann, Open physical systems: from
  {GENERIC} to port-{H}amiltonian systems, in: Proceedings of the 23rd
  International Symposium on Mathematical Theory of Systems and Networks, Hong
  Kong, China, 2018, pp. 204--211.

\bibitem{2020HauschildMarheinekeMehrmannMohringBadlyanReinSchmidt}
\href{https://doi.org/10.1007/978-3-030-53905-4_11}{S.-A. Hauschild,
  N.~Marheineke, V.~Mehrmann, J.~Mohring, A.~M. Badlyan, M.~Rein, M.~Schmidt,
  Port-{H}amiltonian modeling of district heating networks, in: Progress in
  Differential-Algebraic Equations {II}, Springer International Publishing,
  2020, pp. 333--355.
\newblock \path{doi:10.1007/978-3-030-53905-4_11}.}

\bibitem{1983Weinstein}
\href{https://doi.org/10.4310/jdg/1214437787}{A.~Weinstein, The local structure
  of {P}oisson manifolds, Journal of Differential Geometry 18~(3) (1983).
\newblock \path{doi:10.4310/jdg/1214437787}.}

\bibitem{1999MarsdenRatiu}
\href{https://doi.org/10.1007/978-0-387-21792-5}{J.~E. Marsden, T.~S. Ratiu,
  Introduction to Mechanics and Symmetry, Springer New York, 1999.
\newblock \path{doi:10.1007/978-0-387-21792-5}.}

\bibitem{2013Bursztyn}
\href{https://doi.org/10.1017/CBO9781139208642.002}{H.~Bursztyn, A brief
  introduction to {Dirac} manifolds, in: A.~Cardona, I.~Contreras, A.~F.
  Reyes-Lega (Eds.), Geometric and Topological Methods for Quantum Field
  Theory: Proceedings of the 2009 Villa de Leyva Summer School, Cambridge
  University Press, 2013, p. 4–38.
\newblock \path{doi:10.1017/CBO9781139208642.002}.}

\bibitem{2007CerveraSchaftBanos}
\href{https://doi.org/10.1016/j.automatica.2006.08.014}{J.~Cervera, A.~{van der
  Schaft}, A.~Ba{\~n}os, Interconnection of port-{H}amiltonian systems and
  composition of {D}irac structures, Automatica 43~(2) (2007) 212--225.
\newblock \path{doi:10.1016/j.automatica.2006.08.014}.}

\bibitem{2011BatlleMassanaSimo}
\href{https://doi.org/10.1109/cdc.2011.6160588}{C.~Batlle, I.~Massana, E.~Simo,
  Representation of a general composition of {D}irac structures, in: {IEEE}
  Conference on Decision and Control and European Control Conference, {IEEE},
  2011, pp. 5199--5204.
\newblock \path{doi:10.1109/cdc.2011.6160588}.}

\bibitem{1998BlochCrouch}
\href{https://doi.org/10.1090/pspum/064}{A.~M. Bloch, P.~E. Crouch,
  Representations of {D}irac structures on vector spaces and nonlinear {LC}
  circuits, in: G.~Ferreyra, R.~Gardner, H.~Hermes, H.~Sussmann (Eds.),
  Differential Geometry and Control, American Mathematical Society, 1998, p.
  103–117.
\newblock \path{doi:10.1090/pspum/064}.}

\bibitem{2014SchaftJeltsema}
\href{https://doi.org/10.1561/2600000002}{A.~van~der Schaft, D.~Jeltsema,
  Port-{H}amiltonian systems theory: An introductory overview, Foundations and
  Trends in Systems and Control 1~(2) (2014) 173--378.
\newblock \path{doi:10.1561/2600000002}.}

\bibitem{1824Carnot}
S.~Carnot, Réflexions sur la puissance motrice de feu et sur les machines
  propres à développer cette puissance, Bachelier, Paris, 1824.

\bibitem{1848Thomson}
W.~Thomson, On an absolute thermometric scale founded on {C}arnot's theory of
  the motive power of heat, and calculated from {R}egnault's observations,
  Philosophical Magazine (1848).

\bibitem{1854JouleThomson}
\href{https://doi.org/10.1098/rstl.1854.0016}{J.~P. Joule, W.~Thomson, {XV}.
  {O}n the thermal effects of fluids in motion - part {II}, Philosophical
  Transactions of the Royal Society of London 144 (1854) 321--364.
\newblock \path{doi:10.1098/rstl.1854.0016}.}

\bibitem{1850Clausius}
\href{https://doi.org/10.1002/andp.18501550403}{R.~Clausius, {U}eber die
  bewegende {K}raft der {W}ärme und die {G}esetze, welche sich daraus für die
  {W}ärmelehre selbst ableiten lassen, Annalen der Physik und Chemie 155~(4)
  (1850) 500--524.
\newblock \path{doi:10.1002/andp.18501550403}.}

\bibitem{1852Thomson}
\href{https://doi.org/10.1080/14786445208647126}{W.~Thomson, On a universal
  tendency in nature to the dissipation of mechanical energy, The London,
  Edinburgh, and Dublin Philosophical Magazine and Journal of Science 4~(25)
  (1852) 304--306.
\newblock \path{doi:10.1080/14786445208647126}.}

\bibitem{1856Clausius}
\href{https://doi.org/10.1002/andp.18652010702}{R.~Clausius, {U}eber
  verschiedene für die {A}nwendung bequeme {F}ormen der {H}auptgleichungen der
  mechanischen {W}ärmetheorie, Annalen der Physik und Chemie 201~(7) (1865)
  353--400.
\newblock \path{doi:10.1002/andp.18652010702}.}

\bibitem{1995BejanTsatsaronisMoran}
A.~Bejan, G.~Tsatsaronis, M.~J. Moran, Thermal Design and Optimization, Wiley,
  New York, 1996.

\bibitem{2013Wang}
\href{https://doi.org/10.1504/ijex.2013.055077}{L.~S. Wang, Exergy or the
  entropic drive: waste heat and free heat, International Journal of Exergy
  12~(4) (2013) 491.
\newblock \path{doi:10.1504/ijex.2013.055077}.}

\bibitem{2017Wang}
\href{https://doi.org/10.3390/e19020057}{L.-S. Wang, The second law: From
  {C}arnot to {T}homson-{C}lausius, to the theory of exergy, and to the
  entropy-growth potential principle, Entropy 19~(2) (2017) 57.
\newblock \path{doi:10.3390/e19020057}.}

\bibitem{1985Callen}
H.~Callen, Thermodynamics and an Introduction to Thermostatistics, 2nd Edition,
  John Wiley \& Sons Inc, 1985.

\bibitem{1975CurzonAhlborn}
\href{https://doi.org/10.1119/1.10023}{F.~L. Curzon, B.~Ahlborn, Efficiency of
  a {C}arnot engine at maximum power output, American Journal of Physics 43~(1)
  (1975) 22--24.
\newblock \path{doi:10.1119/1.10023}.}

\bibitem{2007Miranda}
\href{https://doi.org/10.7227/IJMEE.35.1.6}{E.~N. Miranda, On the maximum
  efficiency of realistic heat engines, International Journal of Mechanical
  Engineering Education 35~(1) (2007) 76--78.
\newblock \path{doi:10.7227/IJMEE.35.1.6}.}

\end{thebibliography}

\end{document}